\documentclass[prb, twocolumn, superscriptaddress, aps]{revtex4-2}

\usepackage[utf8]{inputenc}
\usepackage[english]{babel}
\usepackage{graphicx}
\usepackage{color}
\usepackage{bbm}
\usepackage{mathtools}
\usepackage{amsthm}
\usepackage{comment}
\usepackage{units}
\usepackage{placeins}
\usepackage{hyperref}
\usepackage{amssymb}
\usepackage[normalem]{ulem}  

\usepackage{tikz}

\newcommand{\pL}{\text{$p_{\rm L}$}}
\newcommand{\pR}{\text{$p_{\rm R}$}}
\newcommand{\pLtwo}{\text{$p^2_{\rm L}$}}
\newcommand{\pRtwo}{\text{$p^2_{\rm R}$}}

\newcommand{\pF}{\text{$p_{\rm F}$}}

\newcommand{\eV}{\text{$eV$}}

\newcommand{\LL}{\text{$\rm L$}}
\newcommand{\RL}{\text{$\rm R$}}
\newcommand{\Transmission}{\mathcal{T}}
\newcommand{\Reflection}{\mathcal{R}}
\newcommand{\Fext}{\text{$F_{\rm ext}$}}

\newcommand{\Pphon}{{\cal P}_\text{ph}}
\newcommand{\Jelec}{J_\text{el}}

\newcommand{\ipcms}{Université de Strasbourg, CNRS, Institut de Physique et Chimie des Matériaux de Strasbourg, UMR 7504, F-67000 Strasbourg, France}
\newcommand{\dipc}{Donostia International Physics Center (DIPC), E-20018, Donostia-San Sebastián, Spain}
\newcommand{\lpmmc}{Univ. Grenoble Alpes, CNRS, LPMMC, 38000 Grenoble, France}

\begin{document}

\title{
Going beyond Landauer scattering theory to describe spatially-resolved \\ non-local heating and cooling in quantum thermoelectrics}

\author{Nico G.\ Leumer}\affiliation{\ipcms}\affiliation{\dipc}
\author{Denis M.\ Basko}\affiliation{\lpmmc}
\author{Rodolfo A.\ Jalabert}\affiliation{\ipcms}\
\author{Dietmar Weinmann}\affiliation{\ipcms}\
\author{Robert S.\ Whitney}\affiliation{\lpmmc}

\begin{abstract}
Spatially-resolved heating and cooling in nanostructures is nowadays measured with various nanoscale thermometry techniques, including scanning thermometry. Yet the most commonly used theory of nanoscale heating and thermoelectricity --- Landauer scattering theory --- is not appropriate to model such measurements. Hence, we analyze a minimal model of spatially-resolved heat transfer between electrons and phonons in simple thermoelectric nanostructures.  This combines Landauer scattering formalism with a Boltzmann equation for transport, revealing the non-locality of Joule heating and Peltier cooling induced by a scatterer in a nanowire. 
The corresponding heating or cooling of the phonons is caused by the voltage drop at the scatterer, but is often maximal at a certain distance from the scatterer. This distance is of the order of the electron-phonon scattering length. 
Scanning thermal microscopy, such as SQUID-on-tip thermometers, should detect this non-locality as phonon hot spots and cold spots, spatially separated from the scatterer.
We provide physical arguments explaining the thermoelectric response of the combined system of wire and scatterer, and in particular, why the resulting heating and cooling is sometimes the opposite to that predicted by the standard Landauer scattering theory.
\end{abstract}

\date{July 26, 2024}
\maketitle

\section{Introduction}
\label{section:intro}

Dissipation is a natural by-product of electrical resistance, leading to Joule heating. 
This conversion of electric power into heat is often an inconvenience, but it is also exploited to produce heat and light (electric heaters and filament lamps). 
At the same time, some materials exhibit thermoelectric effects, that are exploited for applications such as refrigeration using Peltier cooling.  
These effects also exist at the nanoscale, where they are often understood using Landauer scattering theory that links charge transport \cite{landauer1970,Engquist1981Jul,buttiker1986,datta1995,imry2002,jalabert2016}, thermal transport \cite{Engquist1981Jul,Pendry1983Jul} and thermoelectric effects \cite{Sivan1986Jan,Butcher1990Jun,Humphrey2005Mar,
Benenti2017Jun,SciPostPhys2022}, in a way that explains a variety of nanoscale experiments \cite{vanhouten1992,svensson2012lineshape,lee2013,zotti2014,Fast2023Apr}. 
However, there has long been a debate in the literature about {\it where} the dissipation actually occurs at the nanoscale. To this we can also add the question of {\it where} thermoelectric cooling occurs.

\begin{figure}
\centering\includegraphics[width =0.96\columnwidth]{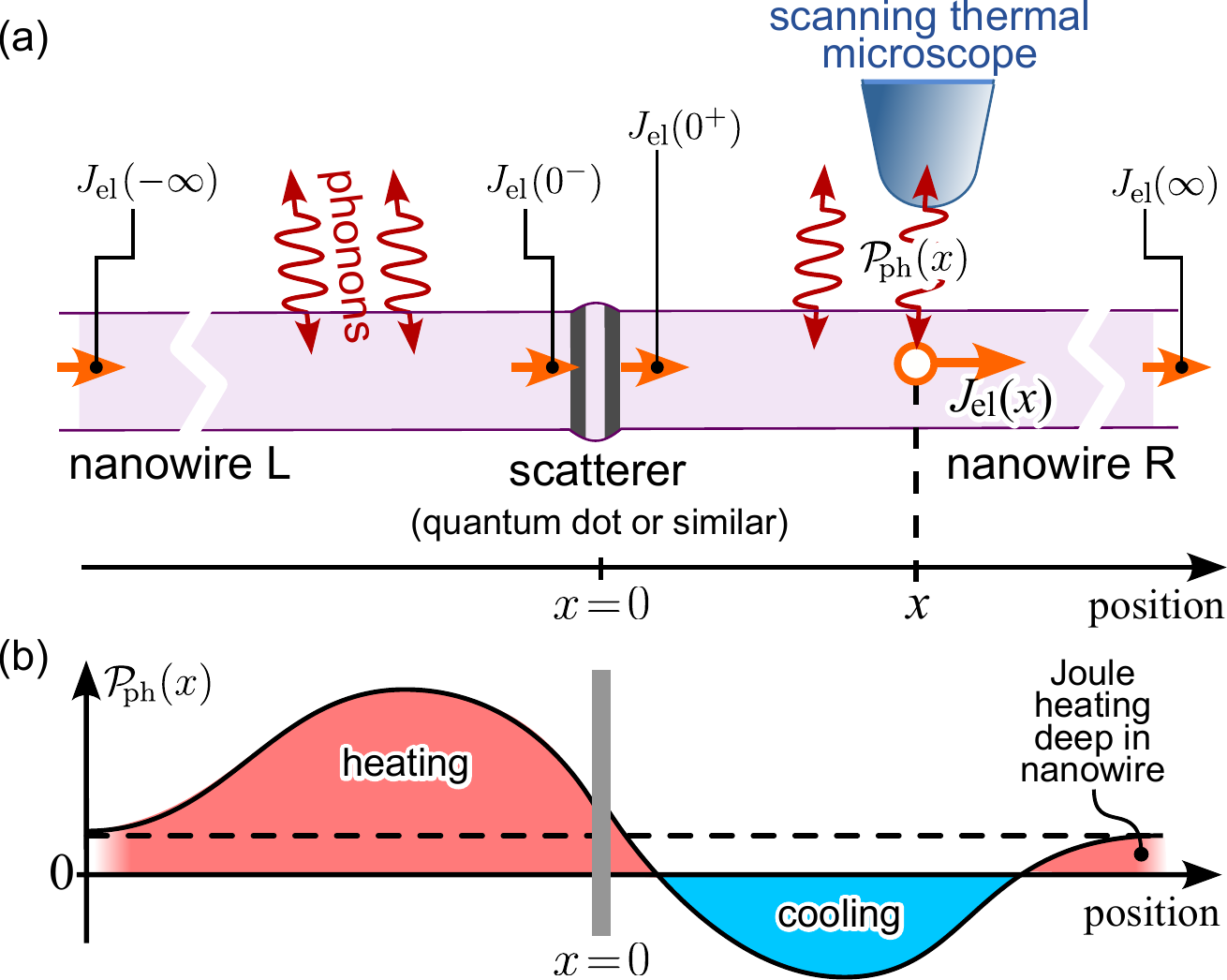}
\caption{
(a) Modeled system of two nanowires (L and R) connected at a scatterer. The electron flow carries a heat current, $\Jelec(x)$ (red arrows), caused by thermoelectric effects and Joule heating.
Heat exchange with phonons 
(heating power per unit length) 
is  $\Pphon(x)$, to be detected at $x$ with a scanning thermal microscope. (b) The electron flow causes phonons  to be heated at some $x$ ($\Pphon(x)>0$) and cooled at other $x$ ($\Pphon(x)<0$).}
\label{Fig_sketch-wire}
\end{figure}

Landauer scattering theory assumes elastic dynamics at the scatterer, so dissipation does not occur at the scatterer (typically a quantum dot, a tunnel barrier, or similar). Instead, all dissipation occurs ``somewhere" in the reservoirs \cite{landauer1970,Engquist1981Jul,buttiker1986,datta1995,imry2002, jalabert2016,Benenti2017Jun,SciPostPhys2022}.
Until recently this statement was sufficient, because we could not study where dissipation occurred experimentally. 
However, the situation has now changed; we now have experimental probes that provide spatial resolution of Joule heating and Peltier cooling at the nanoscale, with a variety of types of scanning thermal microscopy (SThM)\cite{Kim2012May,Gomes2015Mar,Menges2016,halbertal2016,halbertal2017,Weng2018,marguerite2019,zeldov_pc2019}, plasmon resonance microscopy \cite{Mecklenburg2015}, microscopic Raman spectroscopy\cite{Reparaz2014} or spatially-resolved shot-noise probes \cite{Weng2018,weng2021}. 
These probes typically have spatial resolution on the scale of the order of 10--100\,nm, and can resolve temperature differences of order of milli-Kelvins. 
They are providing experimental answers to the question of \emph{where} dissipation occurs. In doing so, they are revealing the richness of non-local dissipation and thermoelectric phenomena. This mandates the development of theoretical models capable of predicting the spatial resolution of heating and cooling at the nanoscale.

In this work, we develop a theory that provides the spatial resolution of the heating and cooling of the phonons (the lattice) caused by electronic currents through nanostructures.  This is relevant to scanning probes (or other local probes) that measure local phonon temperatures, such as in Refs.~\cite{Kim2012May,Menges2016,halbertal2016,halbertal2017,Weng2018,marguerite2019,zeldov_pc2019}.  

Our analysis aims for a minimal model for heating and cooling of phonons, originating from the Joule heating and thermoelectric effects associated with an electronic current through a nanostructure, sketched in Fig.~\ref{Fig_sketch-wire}.
We take a one-dimensional wire containing a single elastic scatterer, where the wire is modeled with a Boltzmann equation whose collision integral corresponds to electron-phonon scattering. 
The electrical current, $I$, induces a Peltier effect that moves heat from one side of the scatter to the other, and a Joule heating that converts electrical work into heat via the dissipative effect of the scatterer's resistance.
Even though we have taken only these minimal ingredients, the model is already complicated: it leads to an infinite set of coupled equations that require numerical solution. As a result, this minimal model has rich physics that was hard to guess from naive arguments, making it worthwhile to understand it before considering more realistic models with additional parameters (such as the finite width of the nanowires, finite electron-electron scattering rates, etc). 
We analyze this minimal model here, showing that it reveals two main physical phenomena, which we then explain.

The first phenomenon that our modeling reveals is that the heating and cooling of phonons to the left and right of the scatterer
are often opposite of what a naive Landauer scattering approach would predict.  
This is not directly related to spatially-resolved heating and cooling effects, but it is necessary to understand them.  This is due to two aspects of the nanowires that are not accounted for in a Landauer approach.  First, the nanowire has a thermoelectric response, and so the nanowire-scatterer-nanowire system should be thought of as three thermoelectrics in series. Second, the electrons in the part of the nanowire near the scatterer exhibit a non-equilibrium distribution, not the simple Fermi distribution assumed in Landauer scattering theory. This electronic distribution is non-equilibrium because neither the scatterer nor the nanowire are in equilibrium when there is a finite current flow.

The second phenomenon is directly related to where dissipation occurs, and is an unexpected consequence of the non-local nature of the heating and cooling of the phonons by the Joule heating and thermoelectric cooling induced by electron flow through the scatterer.  If the heating and cooling of phonons were more local, one would expect that 
heating and cooling would be maximal at the scatterer, and decay away from the scatterer.  
Indeed, this is seen for the hydrodynamic model in Ref.~\cite{Zhang2021Aug} (its Fig.~4 shows that while there may be complicated spatial variations within the scattering region, heating decays smoothly outside that region).
Thus, it is unexpected to find regimes where the maximal heating or cooling of the phonons is at a finite distance from the scatterer (as sketched in Fig.~\ref{Fig_sketch-wire}b), which we will refer to as heating and cooling spots.  Indeed, in some cases, the effect even changes sign as one goes away from the barrier, with phonons being heated close to the scatterer but cooled further way (or vice-versa).  We explain this unexpected phenomenon as follows; when an electron leaves the scatterer and enters the nanowire, that electron flies a certain distance before an electron-phonon scattering event, and this distance depends on that electron's energy. Thus, high-energy electrons  and low-energy electrons thermalize with the phonons at different distances from the scatterer. This means that heating and cooling of phonons can occur at different distances from the scatterer. 

While our model is minimal, and similar models have been treated elsewhere 
\cite{rokni1995,Gurevich1996Jan,eranen1987,laikhtman1994},
we believe the two phenomena that we identify will play a crucial role in local thermometry of nanowires containing scatterers such as a quantum dot.  These could be InAs \cite{svensson2012lineshape,Limpert2017Jul,Josefsson2018Oct,Fast2020Jul,Fast2023Apr} or other types of nanowires, as discussed in more detail in at the end of this paper.  Of course, such experimental nanostructures have a host of parameters that our model does not include (the finite width of the nanowires, inelastic electron-electron scattering, and other scattering mechanisms), some of which are treated in other theoretical models \cite{Zhang2020, Fang2021,Mirlin2019, Tikhonov2018,Zhang2021Aug}.  So we do not expect quantitative agreement between experiments and our minimal model.  Nonetheless, the two phenomena that we identify appear robust enough that it is plausible that they will be present in many nanowires containing quantum dots. Thus, we expect they will be observable in scanning thermal microscopy of suitably chosen nanostructures.

The manuscript is structured as follows. Section \ref{section: model} introduces the model. 
Section \ref{section:results} presents the results showing examples of the two phenomena mentioned above. Section \ref{section:discussion} then discusses and explains these two phenomena. Section \ref{section:experiments} then explains how these phenomena could be observed in experiments.
The basic formalism is presented in Appendix \ref{appendix: Collision integral and Boltzmann}, with some technical details in Appendices,\ref{appendix: matching} and \ref{appendix: I-eV relation}. The formalism is then generalized in Appendix \ref{appendix generic case}.

\section{Model and theoretical approach}
\label{section: model}

\subsection{Choice of the model}

We want a simple model that enables us to study the spatial distribution of energy transfer between electrons and phonons, when there is an electron current that induces both Joule heating and thermoelectric (Peltier) cooling of the electrons. This leads us to consider a one-dimensional geometry: a single-channel wire, interrupted by a scatterer.
For the scatterer, we adopt the well-known Landauer theory of coupled particle and heat current through a scatterer with energy-dependent transmission \cite{Sivan1986Jan,Butcher1990Jun}, as reviewed in ~\cite{Benenti2017Jun}. 
For the wire, the natural framework is the semiclassical Boltzmann's kinetic equation~\cite{Abrikosov2017, DiVentra2008}.
This makes the model similar to those in Refs.~\cite{eranen1987,laikhtman1994}.

We take the wire to be made of two identical semi-infinite wires connected at the scatterer. This provides a simple model for the cases of interest to us where the wire is much longer than the electron-phonon scattering length. 
We assume the scatterer is small enough that electrons pass through it fast enough to not have time to interact with phonons while inside. In other words, electrons enter and leave the scatterer with the same energy, as captured by the Landauer scattering theory.
In contrast, the electron-phonon interactions play a crucial role in the 
wires, and they are modeled as the relaxation in the collision integral of the wires' Boltzmann equations. 
For simplicity, we neglect other relaxation mechanisms in the collision integral.
This is the simplest reasonable description of the physics we investigate here, and will typically describe 
situations where the electron-phonon scattering time is shorter than electron-impurity or electron-electron scattering times.

We consider the steady-state flow, so current conservation means that the electric current $I$ must be the same everywhere in the whole system (in the wire and scatterer). In contrast, the heat current carried by the electrons, $J$, varies along the wire, because electrons are subject to Joule heating and thermoelectric effects, and heat flows between electrons and phonons.
Since the current $I$ and electron-phonon collisions are both present everywhere in the wire, a uniform external electric field must be applied to maintain the constant current. Then, 
the total voltage drop over the infinite wire is infinite, but the current $I$ plays the role of the externally-imposed control parameter.

Coulomb interactions between electrons are included via a self-consistent electrostatic potential. 
This captures the potential buildup in the wire near the scatterer, which we will show to significantly affect the dissipation. We treat Coulomb interactions in the limit of very strong interaction, which then reduces to the requirement of local charge neutrality. We do not include the Coulomb interaction effect on the scatterer's transmission coefficient, since this would require its self-consistent calculation  using some microscopic model for the scatterer. We prefer to avoid this complication and assume the transmission coefficient to be given.

\subsection{Bulk of the wire}
\label{ssec:bulk_wire}

Our model consists of a coherent scatterer placed at $x=0$ between two semi-infinite  1D wires at $-\infty<x<0^-$ and $0^+<x<+\infty$, assumed to be identical. Having in mind doped semiconductor nanostructures, we assume the wires to host one type of charge carriers (either electrons with charge $e<0$, or holes with charge $e>0$). We take the energy dispersion to be parabolic, characterized by the effective mass~$m$ (generalization to a non-parabolic dispersion is straightforward).

In the stationary situation, the Boltzmann equation can be written for each wire as
\begin{align}\label{eq:boltzmann}
	\pm \frac{p}{m} \partial_x f(\pm p, x) \pm \left[\Fext - \partial_x e\varphi(x)\right]\partial_p f(\pm p, x) 
	\nonumber \\  
= \mathcal{I}(\pm p,x) \, ,
\end{align}
where $f(\pm p, x)$ designates the out-of-equilibrium electron distribution function.
The particle momentum is $\pm p$ for right (left) movers (that is, we take $p>0$). $\Fext$~is the external force arising from an applied electric field that drives the current everywhere in the wire, while $\varphi(x)$~represents the electrostatic potential buildup due to the presence of the scatterer, which manifests itself as a local shift of the band bottom by an amount $e\varphi(x)$.

For simplicity, we work in the relaxation-time approximation, where the electron-phonon collision integral in Eq.~\eqref{eq:boltzmann} is written in the form
\begin{align}\label{eq: collisionintegral}
\mathcal{I}(\pm p,x) = -\frac{1}{\tau(p)}\left[f(\pm p, x)-f_T(p,x)\right]\, .
\end{align}
We will concentrate on the case where electron-phonon relaxation time $\tau(p)$ is $p$-independent, while arguing in Appendix~\ref{appendix generic case} that the physics is similar for a generic momentum-dependence of $\tau(p)$. The phonons are assumed to be in good contact with the external bath, so their state is unchanged by the electrons and corresponds to thermal equilibrium with a given uniform temperature~$T$. This is a simplification of the real situation in which the local heating causes small changes in local phonon temperatures. The electron-phonon collisions then tend to relax the electronic distribution to the local equilibrium with the same temperature~$T$:
\begin{align}\label{eq: equilibrium distribution}
f_T(p,x)= \big[\,e^{\beta(p^2/(2m)-\mu[f])}+1\,\big]^{-1} \, ,
\end{align}
where $\beta=(k_\mathrm{B}T)^{-1}$, with $k_\mathrm{B}$ the Boltzmann constant. 
To guarantee the particle conservation by the collision integral, the chemical potential $\mu[f]$ at each point $x$ must be a functional of the non-equilibrium distribution at that point~$x$, determined by the condition
\begin{align} \label{eq: definition mu(x)}
\int_0^\infty \frac{\mathrm{d}p}{h}\, \left[\mathcal{I}(p,x) +  \mathcal{I}(-p,x) \right]=0\, .
\end{align}
The electrostatic potential buildup $\varphi(x)$ should be determined from the 3D Poisson equation involving the excess charge density in the wire, $e[n(x) - n_0]\,\delta(y)\,\delta(z)$, where the 1D particle density,
\begin{align}\label{eq: total density}
n(x) = \int_0^\infty \frac{\mathrm{d}p}{h}\,\left[ f(p,x)+f(-p,x) \right],
\end{align}
and $n_0$ corresponds to the density of the neutralizing background. To avoid solving the Poisson equation, we assume very strong Coulomb interaction, which corresponds to formally sending $e^2\to\infty$ while keeping $e\varphi$ fixed. Then, the Poisson equation can be replaced by the electroneutrality condition,
\begin{align}\label{eq:electroneutrality}
n(x) = n_0.
\end{align}
We emphasize that this represents an equation for $e\varphi(x)$, since the distribution function $f(\pm{p},x)$ found from Eq.~\eqref{eq:boltzmann} depends on the unknown $e\varphi(x)$.

Since $f_\mathrm{T}(p)$ and $e\varphi(x)$ are related to $f(p)$ in a nonlinear fashion, Eq.~\eqref{eq:boltzmann} is a nonlinear integro-differential equation, in spite of its deceptively simple appearance. Thus, we work with its linearized version, assuming that $f(p)$ only weakly deviates from the equilibrium Fermi-Dirac distribution $f_0(p)$.
Here, $f_0(p)$ is the same everywhere in the wires, and is given by Eq.~(\ref{eq: equilibrium distribution}) with $\mu[f]$ replaced by $\mu_0$ chosen such that $f_0(p)$ satisfies~Eq.~\eqref{eq:electroneutrality}.
In other words 
\begin{align}\label{eq: f_0}
f_0(p) = \big[\,e^{\beta(p^2/(2m)-\mu_0)}+1\,\big]^{-1} \, , 
\end{align}
where $\mu_0$ takes the value ensuring that  
\begin{eqnarray}
\label{eq n_0}
    n_0=2\int_0^\infty \frac{\mathrm{d}p}{h}\, f_0(p). 
\end{eqnarray}
This corresponds to $n_0= -\sqrt{2\pi m/\beta} \ {\rm Li}_{1/2}(-e^{\beta\mu_0})$, 
where  ${\rm Li}_n(x)$ is a polylogarithm function.

The parameter that controls the deviation from $f_0(p)$ is the electric current,
\begin{align}\label{eq: current}
I = e \int_0^\infty \frac{\mathrm{d}p}h \, \frac{p}{m} 
\left[f(p, x)-f(-p, x)\right] \, ,
\end{align}
which is the same everywhere in the wires and scatterer in the stationary situation that we consider. 
The heat current in the nanowire is of similar form,
\begin{align} \label{eq:Jelec}
&\Jelec(x) = \int_0^\infty \frac{\mathrm{d}p}{h}\,\frac{p}{m} \left(\frac{p^2}{2m}-\mu_0\right)\left[f(p,x)-f(-p,x)\right],
\end{align}
but as heat is not conserved, this may vary along the nanowire.

Here, we consider the nanowire to linear order in $I$, so we can treat the distribution function to linear order in $\Fext$ and $e\varphi$, since they are both proportional to $I$ for small $I$. 
Thus, we can write the distribution function as the following expansion around the equilibrium distribution, 
\begin{align}\label{eq: distribution function}
	f(\pm p,x) = {}&{}
	 f_0(p) \pm \delta f_{\rm D} (p)
 - e\varphi(x)\,\frac{\partial{f}_0(p)}{\partial\mu_0} + \delta f(\pm p,x)\, ,
\end{align}
The three corrections to the equilibrium $f_0(p)$ here are as follows. The first one,
\begin{align}\label{eq: Drude term}
\delta f_{\rm D} (p) = \Fext\, l(p)\, \frac{\partial{f}_0(p)}{\partial{\mu_0}},\quad 
l(p)\equiv\frac{p}m\,\tau(p),
\end{align}
is the standard Drude contribution that raises (lowers) the right (left) mover occupation, so $f_0(p)\pm\delta f_{\rm D} (p)$ is the translationally-invariant solution for an infinite wire without a scatterer \cite{Abrikosov2017}. In the presence of the scatterer at $x=0$, it is the sole correction in Eq.~\eqref{eq: distribution function} that survives at large distances from the scatterer, $x\to \pm\infty$.
The electrical current at $x=\pm \infty$ can then be calculated from $\delta f_{\rm D} (p)$ alone, giving  
\begin{eqnarray}
I=\sigma\Fext/e.
\end{eqnarray}
This defines the wire conductivity as
\begin{align}\label{eq: wire conductivity}
\sigma \equiv 2e^2\int_0^\infty\frac{\mathrm{d}p}{h}\,\frac{p^2\tau(p)}{m^2}\,\frac{\partial{f}_0(p)}{\partial{\mu_0}}
\ = \ \frac{n_0e^2\tau}m,
\end{align}
where the second equality {\it only} holds for $\tau(p)=\text{const}$.
Similarly, the heat current in the wire carried by the electrons  far from the scatterer ($x\to\pm\infty$) is given by $\delta f_{\rm D} (p)$ alone, with Eq.~\eqref{eq:Jelec} reducing to
\begin{align} 
\label{eq:Jelec infinite x}
&\Jelec(\pm \infty) = 2F_{\rm ext} \int_0^\infty \frac{\mathrm{d}p}{h}\,\frac{p^2\tau(p)}{m^2} \left(\frac{p^2}{2m}-\mu_0\right) \frac{\partial{f}_0(p)}{\partial{\mu_0}},
\end{align}
Here, $\Jelec(\pm \infty)$ is proportional to $F_{\rm ext}$, and hence proportional to $I$, meaning that it is a heat current of thermoelectric origin. It is a Peltier effect that is intrinsic to the wire \cite{Hicks1993May,Mahan1996Jul,Zebarjadi2007Sep}.
It becomes significant when $\mu_0$ in the wire is of order $k_\text{B}T$, so the density of states and electron-phonon scattering length $l(p)= (p/m)\,\tau(p)$ vary significantly over the energy window of order $k_\text{B}T$ around $\mu_0$.

The second correction in Eq.~\eqref{eq: distribution function} arises from the change of the local equilibrium condition as a response to the electrostatic potential buildup $e\varphi (x)$. 
In the strong Coulomb limit, the electroneutrality condition~\eqref{eq:electroneutrality} is satisfied in linear order in $I$ by taking
\begin{align}\label{eq: potential}
	e\varphi(x) =  \left(\frac{\partial{n}_0}{\partial\mu_0}\right)^{-1} \int\limits_0^\infty \frac{\mathrm{d}p}{h} \left[\delta f(p,x)+\delta f(-p,x)\right].
\end{align}

The third correction in Eq.~\eqref{eq: distribution function} includes all other effects.
To get its form, we plug Eq.~\eqref{eq: distribution function} in the kinetic equation~\eqref{eq:boltzmann}, and obtain a \emph{homogeneous} equation for $\delta f(\pm p,x)$.
For this equation, we find a family of linearly independent solutions of the form 
\begin{eqnarray}\label{eq: independent solutions}
g_{l'}(p)[1\mp{l}(p)\,\partial_{x}]e^{\pm{x}/l'}, 
\end{eqnarray}
Each such solution is characterized by $l'$, the distance in $x$ over which it decays, 
and $g_{l'}(p)$ gives the $p$ dependence of that solution.
Using these solutions as a functional basis in the space of functions of~$p$, we expand the unknown $\delta{f}(\pm{p},x)$ in this basis, with the expansion coefficients determined by the boundary conditions at the scatterer, $x=0^\pm$ (see the next subsection).

All technical details of this procedure are provided in Appendix~\ref{appendix: Collision integral and Boltzmann} for the special case of a momentum-independent relaxation time and in the limit of strong Coulomb interaction, while the general case (momentum-dependent relaxation time and arbitrary Coulomb interaction strength) is presented in Appendix\ \ref{appendix generic case}. 
Our conclusion is that the special case of Appendix~\ref{appendix: Collision integral and Boltzmann}, while leading to simpler expressions, is still representative of the general case, so our numerical calculations are performed for this special case.
In contrast, the case $l(p)=\text{const}$, although admitting an exact solution \cite{eranen1987}, turns out to be special and not representative of the general situation.

\subsection{Modeling the scatterer}
\label{ssec:scatterer}

\begin{figure}
\centering\includegraphics[width = 0.7\columnwidth]{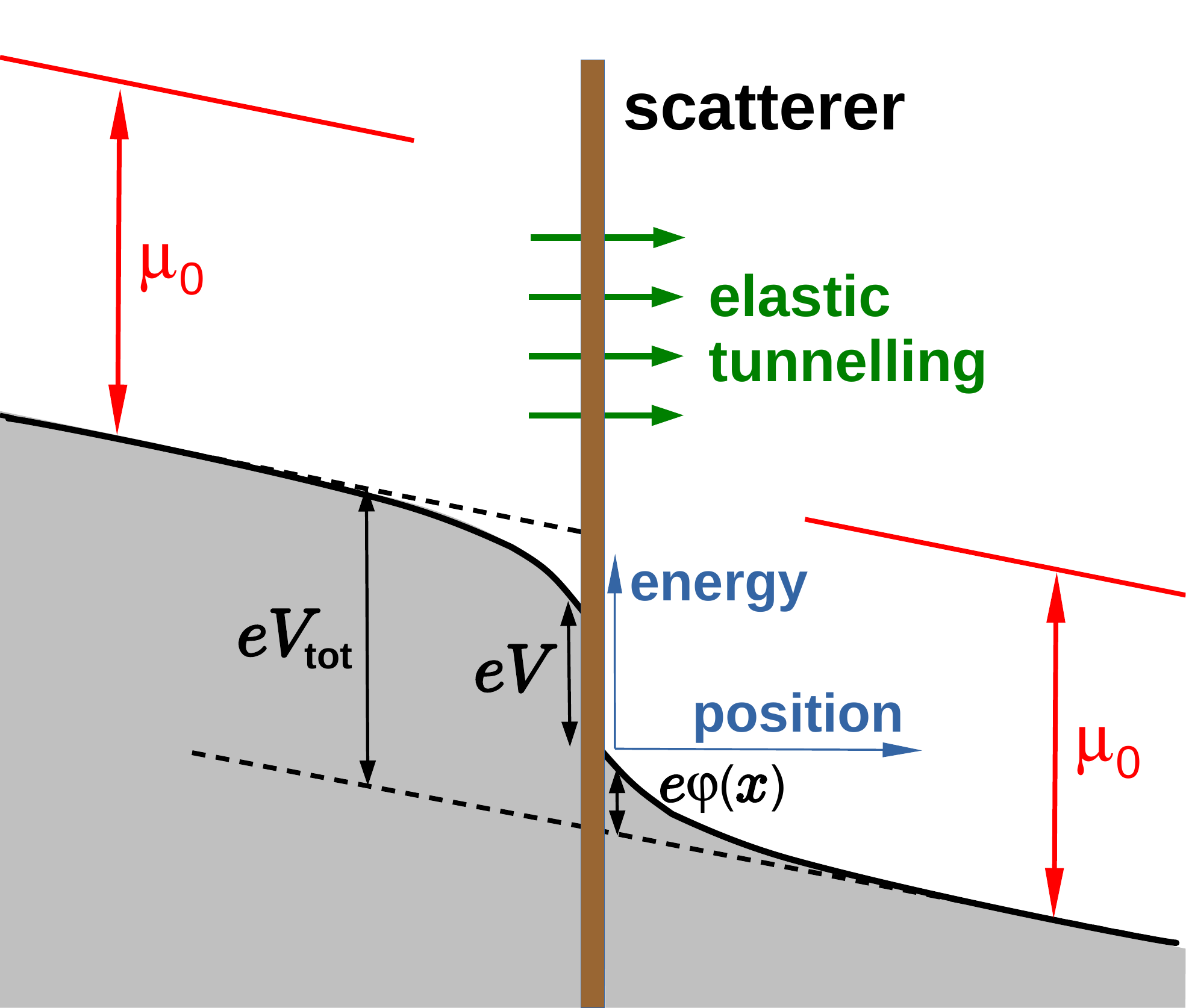}
\caption{Energy diagram in the vicinity of the point scatterer, with $\mu_0$ being the chemical potential far from the scatterer. The electrostatic potential buildup around the scatterer, $e\varphi(x)$, shifts the band bottom in order to accommodate tunneling particles.  Then $V$ is the voltage drop at the scatterer, while $V_{\rm tot}$ is the total voltage drop caused by the presence of the scatterer. }
\label{Fig_sketch_of_potential}
\end{figure}

The presence of a scatterer with a finite reflection probability implies a (yet unknown) voltage drop $V$ which builds up across that scatterer, leading to a misalignment of the bands on the two sides of the scatterer (Fig.~\ref{Fig_sketch_of_potential}). In addition to the linear position-dependence of the band bottom $-\Fext{x}$, imposed by the constant external field $\Fext$, the potential energy buildup $e\varphi(x)$ generates a local modification with respect to the equilibrium case in the region of the scatterer. Therefore, the complete voltage drop induced by the presence of the scatterer is $V_{\rm tot} = V +\varphi(0^+)-\varphi(0^-)$, as illustrated in Fig.~\ref{Fig_sketch_of_potential}.

Since particle transmission is elastic, the change in electrostatic energy $\eV$ upon traversing the scatterer has to be compensated by a change in the kinetic energy (and momentum), so that the total energy~$\epsilon$ is conserved. Namely, a particle incident from the left with momentum~$\pL$ at $x=0^-$ will appear to the right with a different momentum $\pR$ at $x=0^+$, such that 
\begin{align}\label{eq: kinetic energies}
\frac{\pLtwo}{2m} + \eV = \epsilon = \frac{\pRtwo}{2m}\, . 
\end{align}
Thus, the scatterer can be characterized by the transmission probability $\Transmission(\epsilon)$ which depends on the total energy $\epsilon$.
The band misalignment results in full reflection of particles in the right wire approaching the scatterer with kinetic energies below the left band bottom, $0\le\pRtwo/(2m)\le \eV$. Thus, we set $\mathcal{T}(\epsilon<eV)=0$. At higher energies, we have the matching conditions
\begin{subequations}\label{eqs: matching condition}
\begin{align}
f(\pR,0^+) {}&{} = \Transmission(\epsilon)\,f(\pL,0^-) + \Reflection(\epsilon)\,f(-\pR,0^+),\label{eq:matching_right}\\
f(-\pL,0^-) {}&{} = \Reflection(\epsilon)\,f(\pL,0^-) + \Transmission(\epsilon)\,f(-\pR,0^+),\label{eq:matching_left}
\end{align}\end{subequations}
with the reflection coefficient $\Reflection(\epsilon)\equiv1-\Transmission(\epsilon)$, while $\pL,\pR$ are related to~$\epsilon$ via Eq.~(\ref{eq: kinetic energies}). 
Equations~\eqref{eqs: matching condition} automatically satisfy current conservation: we multiply them by $\pR/m$ and by $\pL/m$, respectively, and integrate over momenta taking advantage of $(\pL/m)\,d\pL = d\epsilon = (\pR/m)\,d\pR$, which gives
\begin{subequations}\label{eq:IQPC}\begin{align}
&e\int_{eV}^\infty\frac{\mathrm{d}\epsilon}h\, \Transmission(\epsilon)\left[f(\pL,0^-)-f(-\pR,0^+)\right] \nonumber\\
{}&{} = e\int_0^\infty \frac{\mathrm{d}\pR}h \, \frac{\pR}{m} 
\left[f(\pR, 0^+)-f(-\pR, 0^+)\right],\label{eq:IQPCright}\\
&e\int_{eV}^\infty\frac{\mathrm{d}\epsilon}h\, \Transmission(\epsilon)\left[f(\pL,0^-)-f(-\pR,0^+)\right] \nonumber\\
{}&{} = e\int_0^\infty \frac{\mathrm{d}\pL}h \, \frac{\pL}{m} 
\left[f(\pL, 0^-)-f(-\pL, 0^-)\right].\label{eq:IQPCleft}
\end{align}\end{subequations}
The left-hand side of each equation is nothing but the Landauer scattering theory expression for the current through the scatterer, while the right-hand sides represent the currents in the respective wires near the scatterer, see Eq.~\eqref{eq: current}.

The voltage drop $V$ at the scatterer is not a free parameter of the problem: once we fix the current $I$ (or, equivalently, $\Fext$) at $x\to\pm\infty$, the system must adjust $V$ in order to sustain this current across the scatterer, given its transmission~$\mathcal{T}(\epsilon)$. 
This simple physical argument is not seen explicitly in our equations. Indeed, Eqs.~\eqref{eq:IQPC} follow directly from the matching conditions~\eqref{eqs: matching condition}, and thus do not produce an  independent equation for $V$. Thus, one cannot determine $V$ just by matching the current in the equations. 
In fact, the procedure to find $V$ is more involved. It starts from substituting the ansatz \eqref{eq: distribution function} with Eq.~\eqref{eq: potential} into the matching equations~\eqref{eqs: matching condition}.
The resulting equations for $\delta{f}(p,0^\pm)$ turn out to be incompatible for arbitrary $\Fext$ and~$V$ (see Appendix~\ref{appendix: matching} for details). 
To proceed, one can fix~$V$ and then find the value of $\Fext$ that makes these equations compatible.
Alternatively, one can fix $\Fext$, but then to find the value of $V$ that makes  these equations compatible, one has to go through an implicit self-consistency loop (see Appendix~\ref{appendix: matching}).

Even though we assume small~$I$, we do not necessarily have small $V$ (e.~g., we do not require $I\propto V$). Indeed, if transmission is weak, $\mathcal{T}(\epsilon)\ll1$, or if $\mathcal{T}(\epsilon)\sim1$ only in a very narrow energy interval, a large voltage drop may be required to sustain a small current. One cannot even guarantee that such a value $V$ exists, which is obvious in the extreme limit $\mathcal{T}(\epsilon)=0$.
The non-linear and temperature-dependent $I$-$V$ relation is discussed in Appendix\ \ref{appendix: I-eV relation}, presenting numerical results for a scatterer with an energy-dependent Lorentzian transmission.

While our general approach has no restriction on the form of $\Transmission(\epsilon)$, we particularly study the case of a Lorentzian energy-dependence  
\begin{align}
\label{eq: Lorentzian transmission}
	\Transmission(\epsilon) = \frac{\Gamma^2}{(\epsilon-eV-\epsilon_0)^2+\Gamma^2}
\end{align}
peaked at $\pLtwo/(2m) = \epsilon_0$ and with broadening $\Gamma$.

A scatterer with a Lorentzian transmission window allows for a relatively simple and clear identification of the generic dissipation mechanism, which may be difficult to extract in the case of more general energy dependencies. 
Moreover, a peaked transmission as in Eq.\ \eqref{eq: Lorentzian transmission} can also be relevant for experiments. While QPC's possess a monotonically increasing transmission probability in the incoming particles energy, barriers containing resonant levels can exhibit peaked transmission probabilities. An example are double barrier structures in InAs nanowires with tuneable peak position and broadening $\Gamma$ \cite{svensson2012lineshape}, for which experiments have been performed in a wide range of temperatures.

\subsection{Heating and cooling of phonons}
\label{section: heating/cooling of phonons}

When the distribution function in the wire has been determined, we can calculate the heat flow into the phonons, which is the main subject of our study.
This heating of the phonons at position $x$ is quantified by the heating power per unit length, $\Pphon(x)$, with $\Pphon(x)>0$ for heating of phonons, and $\Pphon(x)<0$ for cooling of phonons. This is defined via the electron-phonon collision integral in Eq.~\eqref{eq:boltzmann},
\begin{equation}
\label{eq: power profile}
   \Pphon(x) = - \int_0^\infty\frac{\mathrm{d}p}{h}\,\frac{p^2}{2m}\, \Big(\mathcal{I}(+p,x) + \mathcal{I}(-p,x)\Big) 
\end{equation}
This can be represented as a sum of 
three contributions~\cite{rokni1995}:
\begin{equation}\label{eq: power profile2}
   \Pphon(x) \ \equiv \ \Pphon^{\rm ext}+\Pphon^{(\varphi)} (x) +\Pphon^{\rm s}(x).
\end{equation}
The first two contributions represent the local Joule heating of phonons at position $x$ in the wire that is due to 
the steady balance between acceleration $\propto \big(F_{\rm ext}-\partial_x e\varphi(x)\big)$ and energy loss of electrons. 
The first of these contributions is for acceleration due to the spatially uniform external field, 
\begin{align}
\label{eq: power through acceleration Fext}
\Pphon^{\rm ext} =\frac{I}{e}\,\Fext   \ \equiv\ \frac{I^2}{\sigma} 
\end{align}
where $\sigma$ is the conductivity of the wire, defined in Eq.~\eqref{eq: wire conductivity}.
The second contribution is for acceleration due to the gradient of the local potential buildup at $x$,
\begin{align}
\label{eq: power through acceleration varphi}
\Pphon^{(\varphi)}(x) =- \frac{I}{e}\, \partial_x e\varphi(x),
\end{align}
which is only significant close to the scatterer.
At small current $I$, Eqs.~\eqref{eq: power through acceleration Fext} and \eqref{eq: power through acceleration varphi}
are both second order, $\mathcal{O}(I^2)$.

The third contribution in Eq.~\eqref{eq: power profile2} is due to the non-equilibrium distribution generated by the scatterer, and reads
\begin{align}
\label{eq: PQPC}
&\Pphon^{\rm s}(x) = -\frac{\partial}{\partial{x}} {\Jelec}(x),
\end{align}
where ${\Jelec}(x)$ is the heat current carried by the electrons at point $x$ in the wire, given in Eq.~\eqref{eq:Jelec}.
This contribution to phonon heating is significant close to the scatterer and vanishes far away from it \footnote{It is convenient to add the $\mu_0$ term in Eq.~\eqref{eq: PQPC}. We are allowed to add this $\mu_0$ term because it cancels, due to it corresponding to $d(\mu_0I)/dx$, and current conservation guaranteeing that $dI/dx$=0}. 
It is linear in $I$ for small $I$, because the electron flow  exhibits a Peltier effect (linear in $I$), and this leads to phonon heating or cooling.

Together $\Pphon^{\rm s}(x)$ and $\Pphon^{(\varphi)}(x)$ represent the phonon heating induced by the scatterer (both Peltier and Joule effects), they are concentrated in the region close to the scatterer, but not precisely at the scatterer's location. Their non-local nature will be discussed in sec.~\ref{sec:non-local}.
In fact, the three contributions to $\Pphon(x)$, discussed above, are nothing but different terms in the continuity equation for the energy density, which follows from the stationary Boltzmann equation~\eqref{eq:boltzmann}:
\begin{equation}
    0 = 
    \left[\frac\Fext{e} - \frac{\partial\varphi(x)}{\partial{x}}\right]I
    - \Pphon(x) -\frac{\partial{\Jelec}(x)}{\partial{x}} .
\end{equation}
where the first term is a source (work done to accelerate  electrons), the second term is a sink (energy going from electrons to phonons), and the last term correspond to the energy transported by electrons.

The contribution $\Pphon^{\rm s}(x)$ can be thought of as the effect on phonon heating of the voltage drop at the scatterer, which causes both a thermoelectric effect and Joule heating.   This is most easily revealed by considering 
the spatial integral of $\Pphon^{\rm s}(x)$  in the left and right half-wires
\begin{subequations}\label{eqs:PLPR}
\begin{align}
  &P^\mathrm{s}_{\LL} \equiv \int_{-\infty}^{0^-}\Pphon^{\rm s}(x)\, \mathrm{d}x  =\Jelec(-\infty)-\Jelec(0^-),\\
  &P^\mathrm{s}_{\RL} \equiv \int_{0^+}^\infty\Pphon^{\rm s}(x)\, \mathrm{d}x = \Jelec(0^+)-\Jelec(+\infty),
\end{align}\end{subequations}
 respectively, where $\Jelec(\pm\infty)$ is given in Eq.~(\ref{eq:Jelec infinite x}). Using the matching conditions~(\ref{eqs: matching condition}), we can rewrite $\Jelec(0^\pm)$ in a Landauer-like form:
\begin{subequations}\label{eqs:energyLandauer}\begin{align}
   \Jelec(0^-) = {}&{} \int\limits_{eV}^\infty\frac{d\epsilon}{h}\, (\epsilon-eV-\mu_0)\nonumber\\
   {}&{}\qquad \times\mathcal{T}(\epsilon)\left[f(p_\LL,0^-) - f(-p_\RL,0^+)\right],\\
   \Jelec(0^+) = {}&{} \int\limits_{eV}^\infty\frac{d\epsilon}{h}\,(\epsilon-\mu_0)\,\mathcal{T}(\epsilon)\left[f(p_\LL,0^-) - f(-p_\RL,0^+)\right].
\end{align}\end{subequations}
where the energy-dependence of $p_{\rm L}$ and $p_{\rm R}$ is given by Eq.~\eqref{eq: kinetic energies}.
In the usual Landauer approach, the distribution functions of the incident particles, $f(p_\LL,0^-)$ and $f(-p_\RL,0^+)$, are assumed to be equilibrium Fermi-Dirac functions, while in our calculation they are found from the Boltzmann equation. In either case, since $\Jelec(-\infty)=\Jelec(+\infty)$, 
we have  
\begin{eqnarray}
P^{\rm s}_{\RL} +P^{\rm s}_{\LL} =  IV.
\end{eqnarray}
So the total phonon heating due to contribution $\Pphon^{\rm s}(x)$ is $IV$, which is what one would expect for the Joule heating due to the voltage drop at the scatterer, $V$.

The contribution $\Pphon^{(\varphi)}(x)$ can be thought of as the effect on phonon heating of the voltage drop due to the potential build up $e\varphi(x)$ induced by the scatterer. 
This is most easily revealed by considering 
the spatial integral of $\Pphon^{(\varphi)}(x)$  in the left and right half-wires,
\begin{subequations}\label{eqs:PLPR-phi}
\begin{align}
  &P^{(\varphi)}_{\LL} = \int_{-\infty}^{0^-}\Pphon^{(\varphi)}(x)\, \mathrm{d}x  = -I\,\varphi(0^-),\\
  &P^{(\varphi)}_{\RL} = \int_{0^+}^\infty\Pphon^{(\varphi)}(x)\, \mathrm{d}x = I\,\varphi(0^+),
\end{align}\end{subequations}
recalling that $\varphi(\pm\infty)=0$.
Taking the sum of the two gives  
$P^{(\varphi)}_{\RL} +P^{(\varphi)}_{\LL} =  I (\varphi(0^+)-\varphi(0^-))$, which is the Joule heating for the voltage drop $ (\varphi(0^+)-\varphi(0^-))$.  Thus, the total Joule heating induced by the
scatterer is 
\begin{eqnarray}
& & \hskip -5mm P^{\rm s}_{\RL} +P^{\rm s}_{\LL} + P^{(\varphi)}_{\RL} +P^{(\varphi)}_{\LL} 
\nonumber \\
&=& I\,\big(V+\varphi(0^+)-\varphi(0^-)\big)  
\ \equiv\  I\,V_{\rm tot},
\label{Eq:total_Joule_heating_due_to_scatterer}
\end{eqnarray}
where $V_{\rm tot}$ is indicated in Fig.~\ref{Fig_sketch_of_potential}.

Finally, we emphasize that all expressions of this subsection are rather general: they do not rely on the specific form of the electron-phonon collision integral or on any assumption about the Coulomb interaction strength.

\section{Numerical Results}
\label{section:results}
\subsection{Emergence of heating and cooling spots}

\begin{figure}[t]
	\centering\includegraphics[width = 0.82\columnwidth]{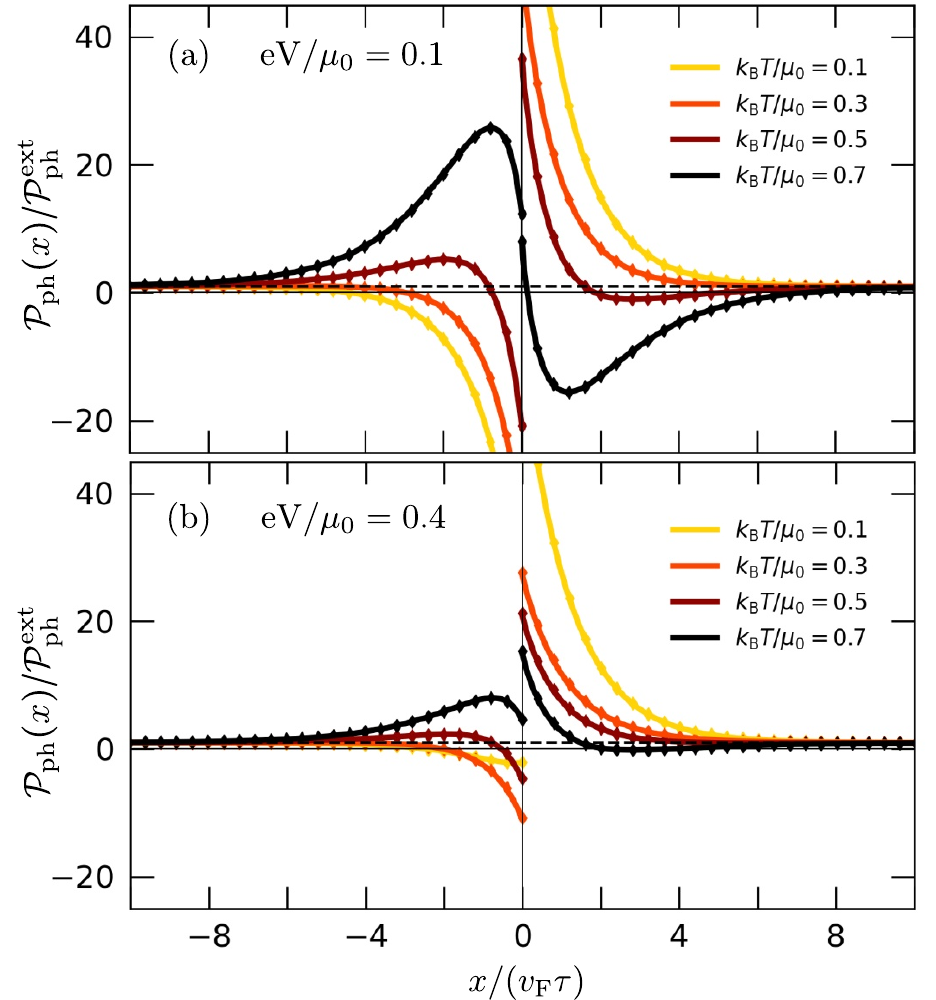}\vspace{-0.2cm}
	\caption{The spatial profile of the heat flow from electrons to phonons, in the vicinity of a scatterer with a Lorentzian transmission of the width $\Gamma/\mu_0 = 0.08$, centered above the left wire chemical potential ($\epsilon_0/\mu_0 = 1.44$) for two values of the voltage drop, (a) $\eV/\mu_0 = 0.1$ and (b) $\eV/\mu_0 = 0.4$. Different colors correspond to different temperatures, $k_\text{B}T/\mu_0=0.1,\,0.3,\,0.5,\,0.7$. Solid lines correspond to heat flow into the phonons, which is plotted in units of the heat flow into the phonons far from the scatterer, $\Pphon^{\rm ext} =\Fext I/e$.  The diamond symbols show $\Pphon(x)-\Pphon^{(\varphi)}(x)$, which is indistinguishable from $\Pphon(x)$ on this scale, and thus $\Pphon^{(\varphi)}(x)$ is negligible for these parameters. The dashed horizontal line shows the asymptote for $x\pm \infty$, which is  $\Pphon(x\to\pm\infty)/\Pphon^{\rm ext}=1$.}
	\label{Fig-spatial-profile}
\end{figure}

\begin{figure}[t]
	\centering\includegraphics[width = 0.82\columnwidth]{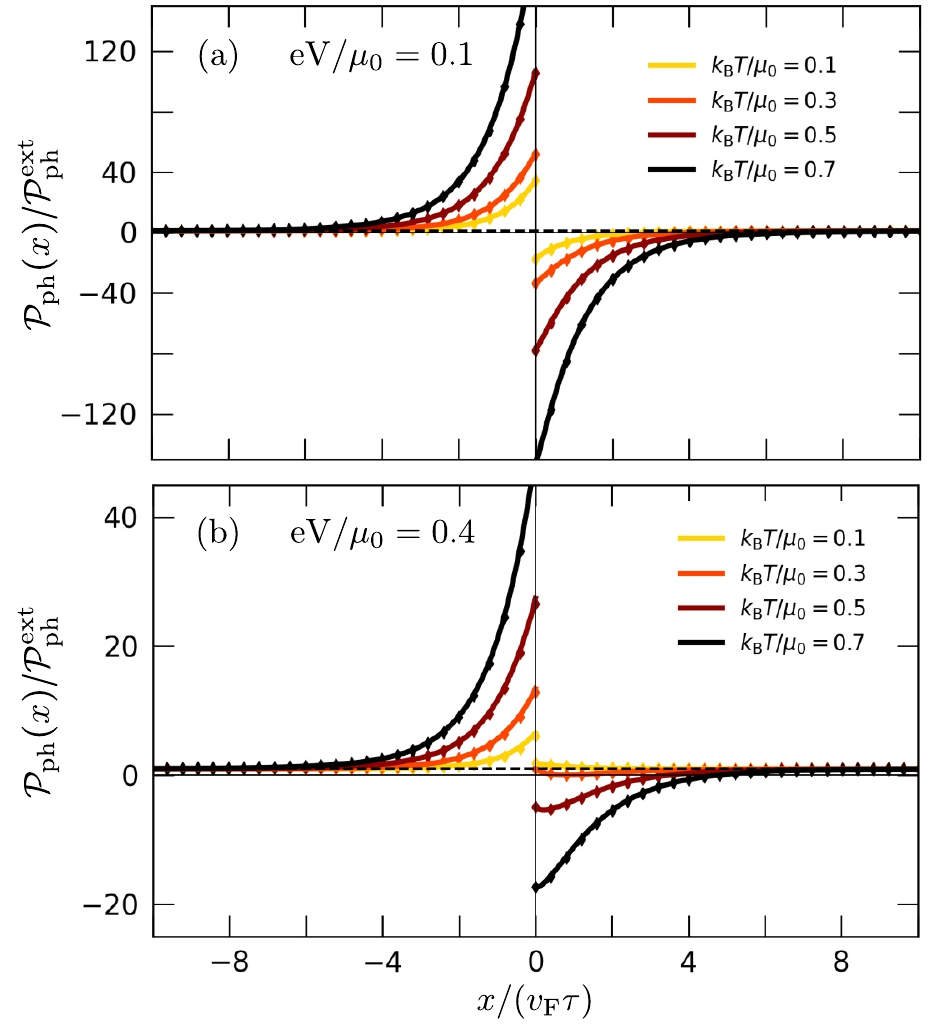}\vspace{-0.2cm}
	\caption{The same as in Fig.~\ref{Fig-spatial-profile}, but for the transmission $\mathcal{T}(\epsilon)$ peaked at $\epsilon_0/\mu_0 = 0.64$, below the left wire chemical potential. 
 }	
	\label{Fig-spatial-profile2}
\end{figure}

Here we present the results for $\Pphon(x)$ in the units of~$\Pphon^{\rm ext}$, obtained numerically by following the procedure described in Sec.~\ref{section: model}. 
Figure \ref{Fig-spatial-profile} corresponds to the case where the scatterer is a quantum dot with a Lorentzian transmission $\mathcal{T}(\epsilon)$ of the form \eqref{eq: Lorentzian transmission} peaked above the chemical potential of the left wire, with $\epsilon_0/\mu_0=1.44$ and $\Gamma/\mu_0 = 0.08$, for two values of the voltage drop~$eV$.
Similar results for the Lorentzian transmission peaked below the chemical potential, namely, at $\epsilon_0/\mu_0 = 0.64$ ($\Gamma/\mu_0 = 0.08$ is the same), are shown in Fig.~\ref{Fig-spatial-profile2}.
The corresponding values of the current $I$ [in the natural units, $Ih/(e\mu_0)$] are given in Table~\ref{tab:currents}.
Based on our expansion in orders of the current,  $\Pphon^{\rm s}(x)$ dominates for small currents, because we saw that $\Pphon^{\rm ext}$ and $\Pphon^{(\varphi)}(x)$ are both $\mathcal{O}(I^2)$, while $\Pphon^{\rm s}(x)\sim \mathcal{O}(I)$
(see Sec.~\ref{section: heating/cooling of phonons} above).
This means that $\Pphon(x)$ near the scatterer can strongly exceed its asymptotic value at $x\to\pm\infty$, since $\Pphon(\pm\infty) = \Pphon^{\rm ext}$.

\begin{table}[]
    \centering
    \begin{tabular}{|c||c|c|c|c|}
    \hline
         & $\frac{\epsilon_0}{\mu_0}=1.44$ &  $\frac{\epsilon_0}{\mu_0}=1.44$ &  $\frac{\epsilon_0}{\mu_0}=0.64$  &  $\frac{\epsilon_0}{\mu_0}=0.64$ \\
         & $\frac{eV}{\mu_0}=0.1$
         & $\frac{eV}{\mu_0}=0.4$
         & $\frac{eV}{\mu_0}=0.1$
         & $\frac{eV}{\mu_0}=0.4$ \\ \hline\hline
         $\frac{k_\text{B}T}{\mu_0}=0.1$ & 0.00531  & 0.0125 & 0.0105 & 0.00832 \\
         $\frac{k_\text{B}T}{\mu_0}=0.3$ & 0.0118 & 0.0362 & 0.036 & 0.0305 \\
         $\frac{k_\text{B}T}{\mu_0}=0.5$ & 0.0166 & 0.0174 & 0.012 & 0.00889 \\
         $\frac{k_\text{B}T}{\mu_0}=0.7$ & 0.158 & 0.08 & 0.051 & 0.037 \\
         \hline
    \end{tabular}
    \caption{The values of the (dimensionless) current $Ih/(e\mu_0)$ corresponding to the different conditions plotted in Figs.~\ref{Fig-spatial-profile} and \ref{Fig-spatial-profile2}.}
    \label{tab:currents}
\end{table}

\begin{table}[]
    \centering
    \begin{tabular}{|c||c|c|c|c|}
    \hline
         & $\frac{\epsilon_0}{\mu_0}=1.44$ &  $\frac{\epsilon_0}{\mu_0}=1.44$ &  $\frac{\epsilon_0}{\mu_0}=0.64$  &  $\frac{\epsilon_0}{\mu_0}=0.64$ \\
         & $\frac{eV}{\mu_0}=0.1$
         & $\frac{eV}{\mu_0}=0.4$
         & $\frac{eV}{\mu_0}=0.1$
         & $\frac{eV}{\mu_0}=0.4$ \\ \hline\hline
         $\frac{k_\text{B}T}{\mu_0}=0.1$ & $-0.00059$ & $-0.0024$ & 0.00051 & $-0.000016$ \\
         $\frac{k_\text{B}T}{\mu_0}=0.3$ & $-0.0019$ & $-0.0095$ & 0.012 & $-0.00010$ \\
         $\frac{k_\text{B}T}{\mu_0}=0.5$ & $-0.0029$ & $-0.0018$ & 0.0034 & 0.00014 \\
         $\frac{k_\text{B}T}{\mu_0}=0.7$ & $-0.054$ & $-0.028$ & 0.12 & 0.0077 \\
         \hline
    \end{tabular}
    \caption{The values of the (dimensionless) electrostatic potential buildup  $e\varphi(x=0^+)/\mu_0$ corresponding to the different conditions plotted in Figs.~\ref{Fig-spatial-profile} and \ref{Fig-spatial-profile2}.}
    \label{tab:potentials}
\end{table}

Figures \ref{Fig-spatial-profile} and \ref{Fig-spatial-profile2} show both $\Pphon(x)$, as defined by Eq.~\eqref{eq: power profile}, and $\Pphon(x)-\Pphon^{(\varphi)}(x)$.
The two are practically indistinguishable, which demonstrates that $\Pphon^{(\varphi)}$ in Eq.~\eqref{eq: power through acceleration varphi} is negligible for the chosen parameters.
Instead of plotting the profile of $e\varphi(x)$, we note that within the approximations used in our numerical calculation (linearized kinetic equation, momentum-independent relaxation time, strong Coulomb interaction), $e\varphi(x)$ has the same spatial dependence as the phonon heating $\Pphon^{\rm s}(x)$ (see the end of Appendix~\ref{appendix: Collision integral and Boltzmann} for the derivation):
\begin{equation}\label{eq:Ps=phi}
   e\varphi(x) = -\frac{2\tau}{n_0}\,\Pphon^\text{s}(x).
\end{equation}
Hence, the spatial dependence of $e\varphi(x)$ is given by the same curves in Figs.~\ref{Fig-spatial-profile} and~\ref{Fig-spatial-profile2}, flipped vertically. 
The magnitude of $e\varphi(x)$ can of course be found from $n_0$ in Eq.~\eqref{eq n_0}, but for convenience we indicate this magnitude
by giving the values of $e\varphi(x=0^+)/\mu_0$ in Table~\ref{tab:potentials}.
We can see from this that $e\varphi(x)$ is small compared to $\mu_0$, $k_{\rm B}T$, as required to justify our linearization of the Boltzmann equation in the wires.

The curves shown in Fig.~\ref{Fig-spatial-profile} present some features that disagree with the common intuition based on the Landauer approach. In particular, the non-monotonic behavior of $\Pphon(x)$ indicates the emergence of heating or cooling spots, which are local maxima or minima of $\Pphon(x)$, respectively. In order to understand their emergence,  we first discuss in Sec.~\ref{sec:asymm} the overall origin of the $\mathcal{O}(I)$ Peltier contribution determining the left-right asymmetry in phonon heating, and then we discuss how a spatially non-monotonic profile arises in Sec.~\ref{sec:discussion:spatial-profile}.

\subsection{Ubiquity of heating and cooling spots}

\begin{figure}[t]
	\includegraphics[width= 0.88\columnwidth]{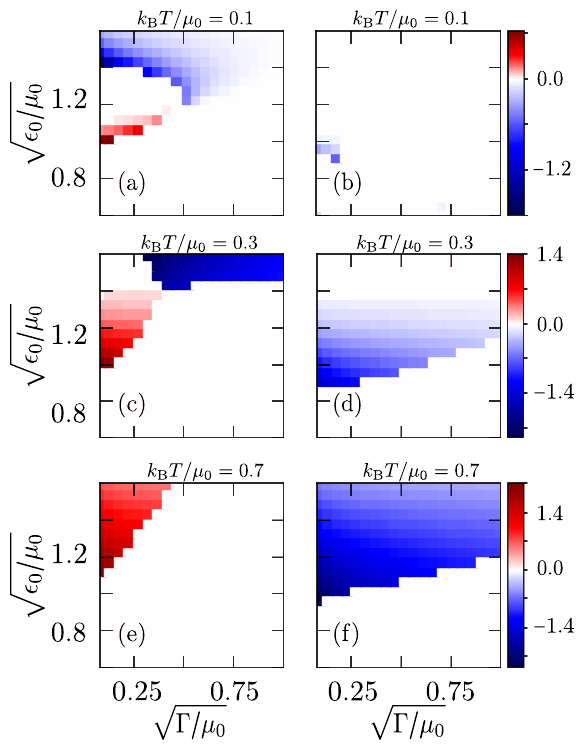}
	\caption{The amplitude of heating spots (positive maximum of $\Pphon(x)$) in red, and the amplitude of cooling spots (negative minimum of $\Pphon(x)$) in blue. The plots are for fixed $\eV/\mu_0 = 0.1$ and different temperatures. The horizontal (vertical) axis indicates the width (position) of the transmission peak, given in Eq.~\eqref{eq: Lorentzian transmission}. The color code shows $\pm[|\Pphon(x_0)|/\Pphon^{\rm ext}-1]^{1/5}$ with red and blue corresponding to heating and cooling spots, respectively. Left (right) column relates to the left (right) wire.}
	\label{Fig-spot-amplitudes}	
\end{figure}

\begin{figure}[t]
	\includegraphics[width= 0.9\columnwidth]{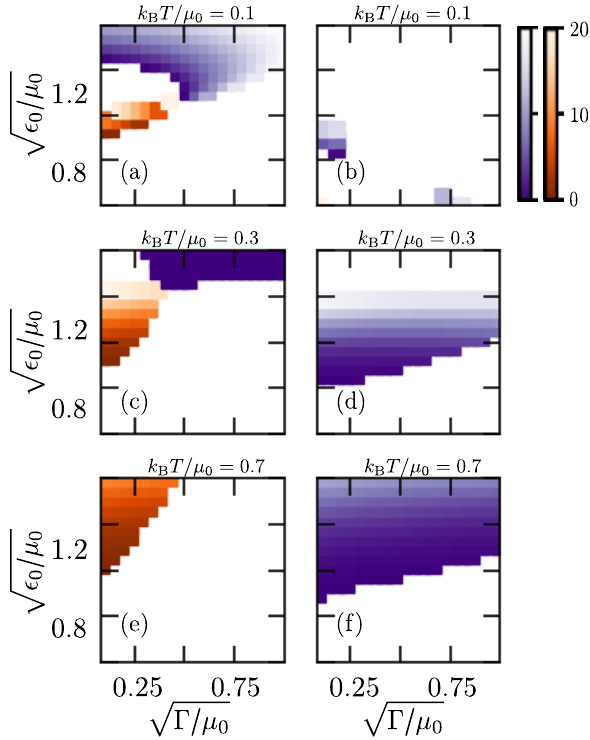}
	\caption{Heating and cooling spot position $\vert x_0\vert/(v_{\rm F}\tau)$ as a function of $\epsilon_0$ and $\Gamma$ for the same parameters as in Fig. \ref{Fig-spot-amplitudes}. Orange/blue  colors correspond to heating and cooling spots. }
	\label{Fig-spot-positions}	
\end{figure}

To check how common heating and cooling spots are, we systematically check for their existence
for Lorentzian transmission, Eq.~\eqref{eq: Lorentzian transmission}, as we vary the Lorentzian's position and width.
The results are given in Figs.~\ref{Fig-spot-amplitudes} and Fig.~\ref{Fig-spot-positions}, which show the amplitude and position of 
local maxima of heating of the phonons (in red) and local minima of heating of the phonons (in blue). 
The color white indicates either that no maxima or minima exist (amplitude of phonon heating decays monotonic with increasing $|x|$), or that the maxima or minima have an amplitude too small to be observable.
For simplicity, Figs.~\ref{Fig-spot-amplitudes} and Fig.~\ref{Fig-spot-positions} only show the dominant maxima or minima, and are restricted to those in the range $|x| < 20\, l(\pF)$ (anyway those at large $|x|$ typically have very small amplitudes, so are unlikely to be observable).
We vary the position of the transmission function's peak $\sqrt{\epsilon_0/\mu_0}$ between $0.6$ and $1.6$, and its width $0.08\,\mu_0<\Gamma<\mu_0$, for fixed $\eV/\mu_0 = 0.1$, and three different temperatures $k_{\rm B}T /\mu_0 = 0.1,\, 0.3,\, 0.7$.

Fig.~\ref{Fig-spot-amplitudes} shows the strength of existing local extrema of $\Pphon(x)$ by color plotting $[|\Pphon(x_0)|/\Pphon^{\rm ext}-1]^{1/5}$ for $x<0$ ($x>0$) in the left (right) column.
Red (blue) color indicates heating (cooling) spots.
The color scale uses a power of 1/5 to better visualize the data (accentuating the color at small values without saturating the color at large values). Comparing panels (c)--(f), we observe that heating (cooling) spots appear in the left (right) wire when the Lorentzian's peak at $\epsilon_0$ is close to the Fermi level. Higher temperatures [panels (e), (f)] support spots at higher $\epsilon_0$, and for larger broadening $\Gamma$. Additionally, we observe a transition from heating and cooling spots as we vary parameters in panels (a) and~(c).

Fig.~\ref{Fig-spot-positions} shows the distance between the scatterer and the point of maximum heating or cooling, $x_0$. 
Comparing  Fig. \ref{Fig-spot-amplitudes} and Fig.~\ref{Fig-spot-positions}, we see a correlation between distance $x_0$ and amplitude of spots. Generally speaking, spots further from the scatterer (larger $x_0$) are smaller in amplitude.

\section{Qualitative picture}
\label{section:discussion}

\subsection{Why the left-right asymmetry strongly differs \\from Landauer scattering theory}
\label{sec:asymm}

\begin{figure}
\includegraphics[width=0.95\columnwidth]{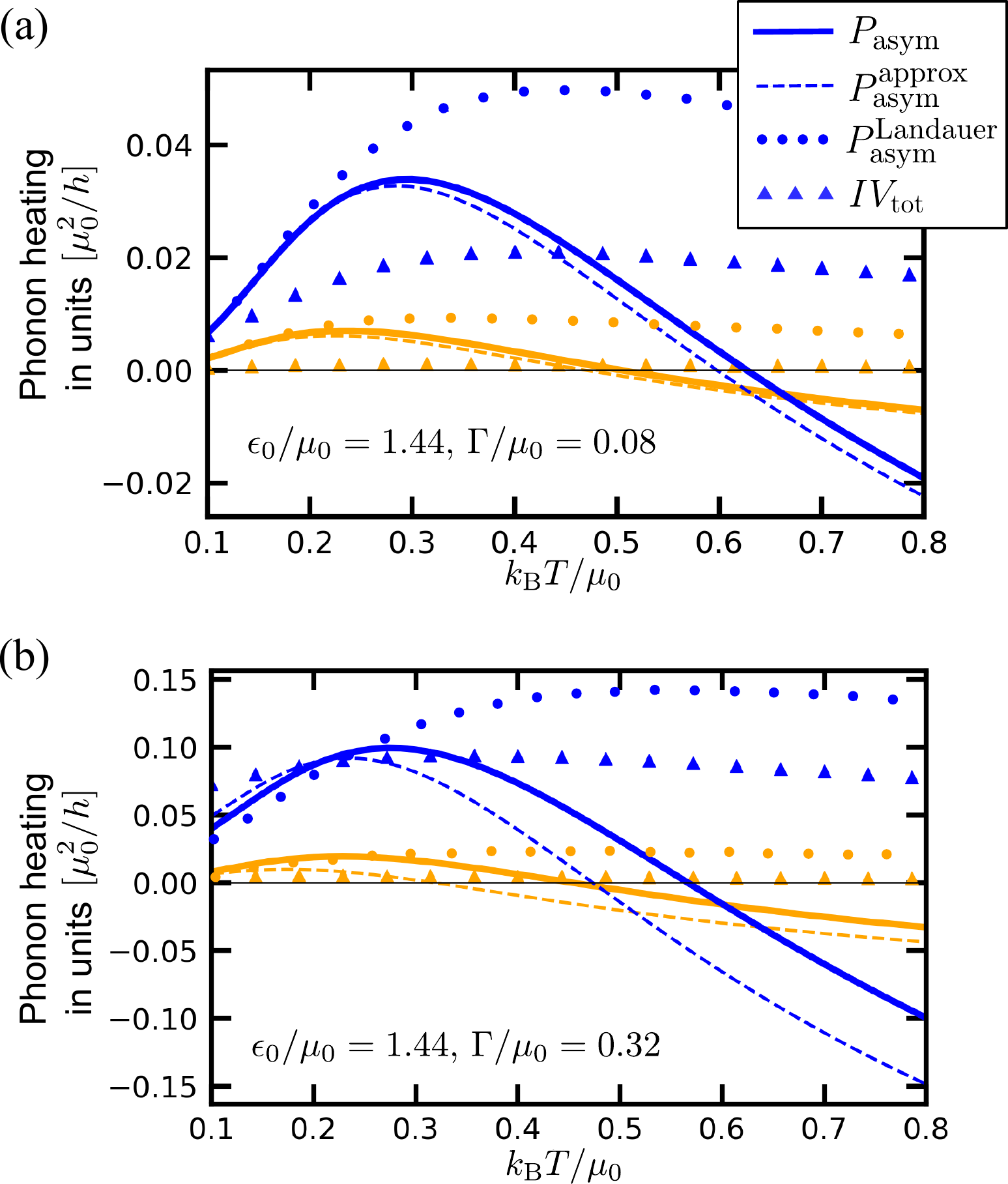}
    \caption{Left-right asymmetry in heating of the phonons versus temperature, for two values of the voltage drop: $eV/\mu_0=0.1,\,0.5$ (orange and blue curves, respectively). 
    In (a) the transmission peak is narrow, $\Gamma/\mu_0=0.08$, while in (b) it is much wider, $\Gamma/\mu_0=0.32$. The solid curves are $P_{\rm asym}$, calculated from our full model. This is very different from the dotted curves, $P_{\rm asym}^{\rm Landauer}$, showing the prediction of Landauer scattering theory in Eq.~\eqref{eq:power_asymmetry-Landauer}. The dashed curves are the approximation $P_{\rm asym}^{\rm approx}$ in Eq.~\eqref{eq:power_asymmetry-approx}.
    This approximation is good for parameters in (a), but shows its limitations for the parameters in (b).
    For comparison, triangles ($\blacktriangle$) show $IV_{\rm tot}$, which is the total contribution to phonon heating (left+right) due to the presence of the scatterer, given by Eq.~\eqref{Eq:total_Joule_heating_due_to_scatterer}. }
    \label{fig:PA_vs_T}
\end{figure}

Before looking at the details of the phonon heating's spatial distribution, 
we focus on the most obvious feature of Fig.~\ref{Fig-spatial-profile}, the asymmetry between phonon heating on the left and on the right of the scatterer. This left-right asymmetry can be quantified by the difference between the integrated  phonon heating/cooling on each side of the scatterer. Using Eqs.~(\ref{eqs:PLPR}) and (\ref{eqs:PLPR-phi}), the asymmetry can be written as
\begin{eqnarray}
\label{eq:power_asymmetry}
    P_\mathrm{asym} &\equiv& 
    P^\mathrm{s}_{\RL}-P^\mathrm{s}_{\LL} +P^{(\varphi)}_{\RL}-P^{(\varphi)}_{\LL} 
    \nonumber \\
    &=& 
    \Jelec(0^+)-\Jelec(+\infty) + \Jelec(0^-)-\Jelec(-\infty) 
    \nonumber \\
    & & {}+ I \left[\varphi(0^-) +\varphi(0^+)\right].
\end{eqnarray}
In contrast, Landauer theory would predict
that $P_\mathrm{asym}$ equals 
\begin{eqnarray}
\label{eq:power_asymmetry-Landauer}
P_\mathrm{asym}^\mathrm{Landauer} = \Jelec^{\rm Landauer}(0^+)+ \Jelec^{\rm Landauer}(0^-),
\end{eqnarray}
with $\Jelec^{\rm Landauer}(0^\pm)$ 
given by Eqs.~\eqref{eqs:energyLandauer} with the distributions replaced by the equilibrium Fermi distribution in Eq.~(\ref{eq: f_0}), i.e., 
$f(p_\LL,0^-)\to f_0(p_\LL)$ and $f(-p_\RL,0^+) \to f_0 (p_\RL)$, where we recall that the $\epsilon$-dependence of $p_{\rm L}$ and $p_{\rm R}$ is in Eq.~\eqref{eq: kinetic energies}.

The solid curves in Fig.~\ref{fig:PA_vs_T} show $P_\mathrm{asym}$, which we see is strongly different 
from the Landauer scattering theory prediction $P_\mathrm{asym}^{\rm Landauer}$ (dotted curves).
Indeed, it often has the opposite sign from the Landauer scattering theory at higher temperatures. 
This may seem surprising, given how Landauer scattering theory is widely used in descriptions of thermal and thermoelectric effects in nanostructures.  So it is important to identify where the difference comes from.
The reason for the difference is that nanowires are  not the \textit{ideal leads} that are assumed in Landauer scattering theory.
This results in two main differences from Landauer scattering theory.

The first difference between the full solution and the Landauer scattering theory prediction is physically straight-forward; the  nanowires have a thermoelectric response, 
$\Jelec(\pm\infty)$ given by Eq.~\eqref{eq:Jelec infinite x}.
This is absent in Landauer scattering theory, since such a Landauer scattering theory assumes leads with $\mu_0 \gg k_\text{B}T$.
Thus, a nanowire containing a scatterer (as in Fig.~\ref{Fig_sketch-wire}) is actually \textit{three thermoelectrics in series}. The scatterer is a thermoelectric that is sandwiched between two (semi-infinite) nanowires that are also thermoelectrics. The additive effect of the three thermoelectrics can be very different from that of the scatterer alone, particularly when the scatterer's thermoelectric response is smaller than that of the two nanowires, so the thermoelectric response is dominated by the nanowires.
With this in mind we make a simple improvement to Landauer scattering theory,
we assume  $\Jelec^{\rm Landauer}(0^+)$ approximates the heat current leaving the scatterer at $x=0^+$, but then 
note that electrons deep in the wire (at $x\to \infty$) carry heat $\Jelec(\infty)$, so the heat flow into the phonons must be $\Jelec^{\rm Landauer}(0^+)-\Jelec(\infty)$. Making the same argument for the left side of the nanowire, we would predict $P_\mathrm{asym}$ would be approximately 
\begin{eqnarray}
\label{eq:power_asymmetry-approx}
P_{\rm asym}^{\rm approx} = P_{\rm asym}^{\rm Landauer} -\Jelec(+\infty)-\Jelec(-\infty).
\end{eqnarray}
This approximation gives the dashed curves in Fig.~\ref{fig:PA_vs_T}, which gets close to the true result (solid curves) in Fig.~\ref{fig:PA_vs_T}a.

Before discussing why this approximation is good in Fig.~\ref{fig:PA_vs_T}a but not Fig.~\ref{fig:PA_vs_T}b, we note that it allows us to understand why the Landauer scattering theory works better at low temperatures, and why the asymmetry changes sign as $T$ increases. The nanowires' thermoelectric response (for given $I$) is always negligible at small $T$, and then grows with increasing $T$, as seen from Eq.~\eqref{eq:Jelec infinite x}. In contrast, the scatterer's thermoelectric response (for given $I$) may be fairly weakly $T$ dependent.
In that case, it is normal that small $T$ corresponds to a thermoelectric effect dominated by the scatterer, but large $T$ corresponds to one dominated by the nanowires.
Hence, if $P_{\rm asym}$ is positive at small $T$ (when it is dominated by the scatterer), one can guess that it will become negative whenever $T$ becomes large enough that the nanowire's thermoelectric response starts to dominate.

It then becomes easy to understand why the full theory --- or the above approximation of it --- can predict cooling of the phonons on the right of the scatterer, when Landauer predicts heating of those phonons (and vice versa for phonons on the left of the scatterer). A specific example of this is large $T$ in  Fig.~\ref{fig:PA_vs_T}a.  
We already know that the nanowire's Peltier effect induces a heat current in the same direction as the particle current, Eq.~\eqref{eq:Jelec infinite x}.  For simplicity, let us assume the scatterer's Peltier effect is of the same nature
(typical of transmission that is peaked above $\mu_0$, as in Fig.~\ref{fig:PA_vs_T}). 
Thus, when there is a particle current through all three thermoelectrics in series, their electrons all carry heat from left to right (cooling phonons to their left and heating phonons to their right).  However, if the 
middle thermoelectric (the scatterer) exhibits a weaker Peltier effect than the nanowire to its right, then it will be heating phonons to its right by less than the right-hand thermoelectric (the right nanowire) is cooling those phonons.  Thus, the net effect is a cooling of the phonons to the right of the middle thermoelectric, the opposite of what one would think when considering the middle thermoelectric in isolation.
Landauer scattering theory is effectively treating the middle thermoelectric in isolation, and forgetting the  thermoelectric responses of the nanowires.  This can only be justified when $T$ is low enough that the nanowires have a negligible thermoelectric responses.
The approximation in Eq.~\eqref{eq:power_asymmetry-approx} does much better by accounting for the nanowires's thermoelectric responses.

Now we turn to the second main difference between the full solution and the Landauer scattering theory prediction, that is not captured by the approximation in Eq.~\eqref{eq:power_asymmetry-approx}.
In Landauer scattering theory, it is assumed that each lead is macroscopic, and so its electronic distribution is not perturbed by the scatterer or the current flowing through it (for example, the lead conductances are assumed to be infinitely greater than the scatterer's conductance), which is why it takes the distribution in the leads to be $f_0$ from Eq.~\eqref{eq: f_0}.  Here, in contrast, the leads are nanowires with finite conductivities, so their electronic distributions are modified whenever they carry current. Each nanowire's distribution of left and right movers is shifted in opposite directions, given by $\delta f_{\rm D}(p)$ in Eq.~\eqref{eq: Drude term}. Since $\delta f_{\rm D} (p)\propto l(p)\,\partial_{\mu_0}f_0(p)$,  in addition to the thermal broadening, its peak shifts towards higher energies \emph{above} the Fermi level by an amount $\sim{k}_\text{B}^2{T}^2/\mu_0$.
The finite conductance of the nanowires also means there is a potential buildup, $\varphi(x)$, in the nanowire close to the  scatterer,
which not only gives the $P^{(\varphi)}_{\RL}-P^{(\varphi)}_{\LL}$ term in Eq.~\eqref{eq:power_asymmetry},
it also further modifies the electronic distributions that enter $\Jelec(0^\pm)$.   
Unfortunately, these factors are hard to guess and probably require the full self-consistent solution of the problem.
However, in the weak transmission limit, $\mathcal{T}(\epsilon)\ll{1}$, these factors are of order $\mathcal{O}(\mathcal{T}^2)$, since $\delta{f}_D(p)$ is $\mathcal{O}(\mathcal{T})$ with respect to $f_0(p)$, and $P^{(\varphi)} \sim  \mathcal{O}(I \varphi(x)) \sim \mathcal{O}(\mathcal{T}^2)$.  Thus, the approximation $P_{\rm asym}^{\rm approx}$ captures all effects up to $\mathcal{O}(\mathcal{T})$, so it will become exact in the limit of small $\mathcal{T}$.

In a more handwaving manner, we can say that the approximation in Eq.~\eqref{eq:power_asymmetry-approx} is expected to work well when 
the scatterer's conductance is small compared to the conductance quantum. We know it works well when the scatterer's transmission is small at all momenta, but this handwaving argument suggest that it will also capture the physics if the transmission is only large in a very narrow range of momenta, such as the narrow Lorentzian transmission typical of a quantum dot. This would explain why Fig.~\ref{fig:PA_vs_T}a shows the approximation (dashed curve) being close to the full result (solid curve) for the narrow transmission peak, but that this is not the case for the broad transmission peak in Fig.~\ref{fig:PA_vs_T}b.

\subsection{Non-locality of the phonon heating}
\label{sec:non-local}

The phonon heating is clearly not happening locally at the scatterer; the scatterer is at $x=0$ and the heating or cooling of the phonons is often peaked at finite $x$.  However,  we should note that the voltage drop is not happening locally at the scatterer either. While $V$ is the voltage drop {\it at} the scatterer, $V_{\rm tot}$ is the total voltage drop induced {\it by} the scatterer (see Fig.~\ref{Fig_sketch_of_potential}), and it includes the effect of the potential due to the buildup of charges around the scatterer, given by $e\varphi(x)$.

Thus, to be rigorous, it is wise to separate heating at point $x$ into the part that is due to the voltage drop at $x$, and the part that is {\it really non-local}, because it is due to a voltage drop elsewhere.  This separation is easily done. $\Pphon^{(\varphi)}(x)$ is local, because its value at $x$ depends on $\partial_x e\varphi(x)$. In contrast, $\Pphon^{\rm s}(x)$ is non-local because it is due to the voltage drop at the scatterer (at $x=0$).  
They have very different magnitudes for small currents, $I$. The non-local term $\Pphon^{\rm s}(x)$ is $\mathcal{O}[I]$, while the local term $\Pphon^{(\varphi)}(x)$ is $\mathcal{O}[I^2]$. Thus, for small currents,  the heating and cooling at $x \neq 0$ is entirely a consequence of the non-local part of the phonon heating. This is the case in Figs.~\ref{Fig-spatial-profile} and \ref{Fig-spatial-profile2}.

Surprisingly, 
this non-local heating (or cooling) is maximal at a 
finite distance from the scatterer, leading to clear heating and cooling spots.  One would be more likely to expect a monotonic decay of this non-local heating as one moves away from the scatterer. Our next aim is to understand why this expectation is
wrong.

\subsection{Why the spatial profile has heating and cooling spots}
\label{sec:discussion:spatial-profile}

\begin{figure}
    \centering
    \includegraphics[width=0.95\columnwidth]{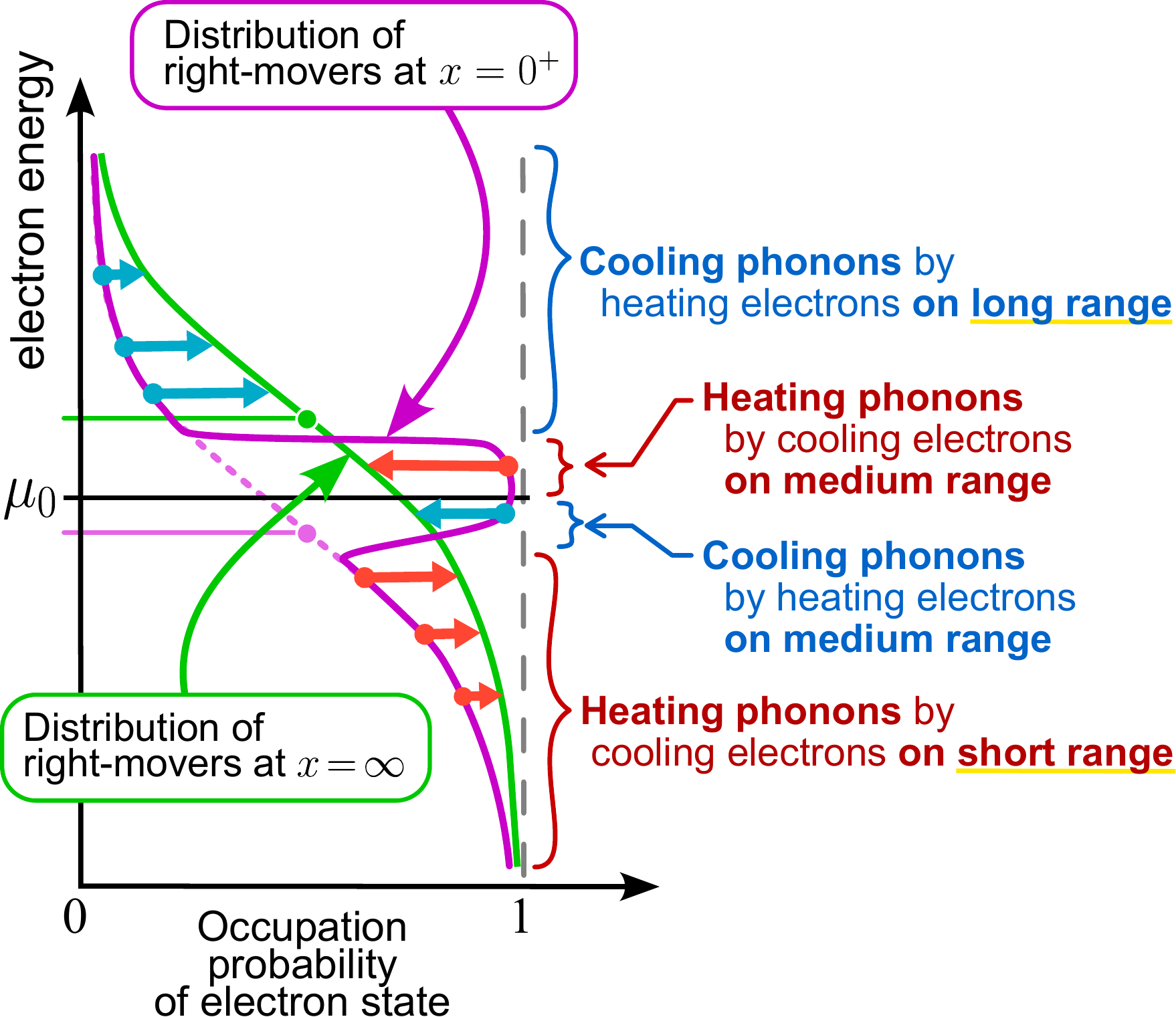}
    \caption{
    A sketch to qualitatively describe the origin of the main features of the spatially-resolved heating/cooling of phonons in Fig.~\ref{Fig-spatial-profile} (specifically the black curve in Fig.~\ref{Fig-spatial-profile}a).
    It shows how the distribution of electrons moving away from the scatterer changes with increasing $x$. It can be thought of as a qualitative interpretation of the term containing $f(p,x)$ that we get from substituting Eq.~(\ref{eq:Jelec}) into Eq.~(\ref{eq: PQPC}), as explained in Sec~\ref{sec:discussion:spatial-profile}. 
    \label{fig:sketch-explaining}}
\end{figure}

Our next goal is to understand the origin of the heating and cooling spots, that is, local maxima and minima of $\Pphon(x)$ at some distance away from the scatterer. 
We start from Eq.~(\ref{eq: power profile}), which tells us the phonon heating depends on the collision integral, $\mathcal{I}(\pm p,x)$, and we recall from Eq.~(\ref{eq: collisionintegral}) and Sec.~\ref{ssec:bulk_wire} that this collision integral is made from eigenfunctions with exponential spatial dependence. Hence, we
conclude that $\Pphon(x)$ is a linear combination of many spatially decaying exponentials with different mean free paths. The key idea can be understood by considering just two decaying exponentials for the case $x>0$, such as the function 
\begin{equation}\label{eq: f(x)}
   \mathcal{F}(x) = A_1e^{-x/l_1} + A_2e^{-x/l_2},\quad \mbox{ with } l_2>l_1.
\end{equation}
This function can have a maximum or a minimum at $x>0$ if and only if (i)~$A_1$ and $A_2$ have different signs, and (ii)~$|A_1| > (l_1/l_2)|A_2|$. For  the maximum/minimum to be sufficiently prominent, $l_1$ and $l_2$ should be neither too close to each other, nor strongly different.
Hence, for Eqs.~(\ref{eq: PQPC}) to produce a heating or cooling spot, we need to identify non-equilibrium contributions to the distribution function at different momenta (with sufficiently different mean free paths), which have different signs. 

As an example, we can consider the sketch in Fig.~\ref{fig:sketch-explaining}, whose aim is to describe the main features of the spatially-resolved heating/cooling of phonons in Fig.~\ref{Fig-spatial-profile}'s black curve.  This sketch is an interpretation of the term containing $f(p,x)$ in Eq.~(\ref{eq:Jelec}) which enters Eq.~(\ref{eq: PQPC}). Before explaining this sketch's construction in detail, one immediately sees that it has blue arrows indicating contributions of one sign (cooling phonons) and red arrows indicating contributions of the other sign (heating phonons), so it has the above ingredient for heating and cooling spots.

The sketch in Fig.~\ref{fig:sketch-explaining} is constructed for Eqs.~(\ref{eq:Jelec}),~(\ref{eq: PQPC}) as follows.
The violet curve is the distribution of right-moving electrons leaving the scatterer at $x=0^+$. This distribution is made of a shifted Fermi function (representing reflection of left-moving electrons from the scatterer) with a peak super-imposed on it (representing right-moving electrons that were transmitted through the scatterer). 
As the right-moving electrons move away from the scatterer, the violet distribution relaxes towards the green distribution by exchanging energy with the phonons. This results in electrons at some energies heating phonons, and electrons at other energies cooling phonons (as indicated).  However, when electrons have more energy, they move faster, and the heat exchange with phonons occurs over a longer distance. Thus, heating and cooling effects happen on different length scales. In the example sketched here, the heating of phonons is dominated by slow (low energy) electrons, while cooling of phonons is dominated by fast (high energy) electrons. Hence, heating happens closer to the scatterer than cooling, as seen clearly to the right of the scatterer in Fig.~\ref{Fig-spatial-profile}a's black curve.  

In this example, where the transmission mostly happens at energies close to $\mu_0$, the tails of the two distributions are due to the finite temperature in the wires. As temperature goes down, these tails shrink, so we lose the long and short range contributions to heating/cooling, which means the heating/cooling spots tend to disappear, as seen in Fig.~\ref{Fig-spatial-profile}a. 

A sketch similar to Fig.~\ref{fig:sketch-explaining} can be made for the left-moving electrons leaving the scatterer, and it captures the main features of the heating of phonons to the left of the scatterer in Fig.~\ref{Fig-spatial-profile}a's black curve. 

Obviously, such sketches cannot be used for quantitative predictions, because they miss the self-consistency conditions that must be satisfied by the distributions. For example, current conservation should self-consistently determine the green distribution's vertical position with respect to the violet distribution, while electro-neutrality should self-consistently determine the band-bending (finite $\varphi(x)$ near the scatterer) that modifies the vertical position of distributions of electrons arriving at the scatterer, as well as those leaving the scatterer.
These self-consistency conditions are absent in this sketch, but they are present in our full theory, and they make our analysis and numerical modeling so involved.

\section{Conditions for experimental observability of hot and cold spots}
\label{section:experiments}

While our model is chosen for its simplicity, it does give a strong indication of conditions for observing phonon hot and cold spots in semiconductor nanowires that contain a quantum dot.
Our model is expected to be a reasonable description of such experiments under the following conditions.
\begin{itemize}
\item  Electron energy relaxation is dominated by electron-phonon collisions, rather than inelastic electron-electron collisions (this is the opposite limit from Refs.~\cite{Tikhonov2018,Zhang2021Aug}).
\item Electrons traverse the scatterer (quantum dot or similar) fast enough that they have no appreciable interaction with phonons while inside that scatterer. 
\item If the nanowire has finite width, then it may have multiple transverse modes. In that case, we require that all transverse modes in the nanowire have similar dynamics (similar scattering at the quantum dot, and similar electron-phonon scattering in the nanowire). Then this case will be qualitatively described by taking our results for a single transverse mode and multiplying by the number of transverse modes.  
\end{itemize}
Then our model shows that heating and cooling spots can be expected
when the semiconducting nanowire has relatively small chemical potential $\mu_0$ (so $k_{\rm B}T$ or $eV$ can be of order $\mu_0$), ensuring that the electron-phonon scattering length varies significantly over the window of electron energies involved in transport (the window of order $k_{\rm B}T$ or $eV$ around $\mu_0$).

The exact spatial profile of the heating will of course depend on the momentum dependence of the electron-phonon scattering length, and the strength of the Coulomb interaction. 
While our numerical results were obtained for a particular choice of these, our minimal model can be set-up for any such momentum dependence and any screening (as shown in Appendix~\ref{appendix generic case}), and the phenomena we describe would generically be present in any of them.

The formulated conditions could be satisfied by samples such as InAs nanowires containing quantum dots, like those whose global thermoelectric response was studied at low temperatures in Ref.\cite{svensson2012lineshape,Josefsson2018Oct,Fast2023Apr},
and at high temperatures in Refs.~\cite{Limpert2017Jul,Fast2020Jul,Fast2022Jun}.  Similar materials appear to have electron energy relaxation dominated by electron-phonon interactions above about 10\,K \cite{Prasad2003Jul}. These nanowires' electrochemical potential, $\mu_0$, can be tuned with doping or electro-static gates, and experimentally $\mu_0$ (as measured from the band-edge) is often of similar order to temperature, thus  we take such parameters in our modeling. It  means that the nanowire itself has a significant thermoelectric response (as observed experimentally in Ref.~\cite{svensson2012lineshape}).

To see if the heating and cooling spots could be resolved by scanning thermal microscopy (SThM), we should be aware that scanning thermal microscopy detects {\it hot spots} (local temperature increase of the phonons), not {\it heating spots} (local heat flow from electrons to phonons, as plotted in our figures). We can expect that the spatial variation of the phonon temperature will correspond to a broadened version of the spatial distribution of heating and cooling spots.  This broadening will be of the order of the phonon mean free path. Thus, for the heating and cooling spots to be clearly visible in the phonon temperature profile, we need that the peaks of the spots are at a distance from the scatterer that is more than a phonon mean free path. 
Our results show that their distance from the scatterer is of order the electron-phonon scattering length.
For semiconducting nanowires at 300\,K this scattering length is typically hundreds of nanometers \cite{Fast2021Jun}, while at low temperatures it is typically longer than at 300\,K. 
In contrast, the phonon mean free path will typically be of order the wire diameter, which can be made 50\,nm or smaller.
In such cases, the heating and cooling spots will be visible despite the smoothing of 
the spatial profile of the phonon temperature.

Finally, one needs to check that this phonon temperature profile (showing hot and cold spots) can be spatially resolved by scanning thermal microscopy. As we already stated our model shows this profile is on the scale of the electron-phonon scattering length, which can be hundreds of nanometers.
Typically, scanning thermal microscopy has a resolution down to about 50\,nm, so there should be no problem in it resolving the hot and cold spots.

\section{Conclusions}
\label{sec: discussion}

This work presented a minimal model to address the question of {\it where} dissipation and thermoelectric effects occur in nanostructures.  It models electron flow through a scatterer (quantum dot or similar) in a nanowire. The nanowire's electrons relax via electron-phonon interactions, modeled with a one-dimensional Boltzmann equation in the relaxation-time approximation.  Despite its apparent simplicity, the model led to an infinite set of coupled equations that require numerical solution.  This resulted in rich physics, exhibiting strong deviations from a more conventional Landauer scattering theory, and unexpectedly non-local phonon heating and cooling spots. These spots are heating (or cooling) of phonons, caused by the Joule heating and thermoelectric effects due to the voltage drop at the scatterer. Yet, these spots are peaked at a finite distance from the scatterer, meaning that the phonon heating and cooling is a non-local effect.  We argued that these heating and cooling spots should be observable with scanning thermal thermometry, such as SQUID-on-tip thermometers.
We also explained many of our results with simple physical arguments, even if our full numerical calculation was necessary to get quantitative results.

The numerical approach presented here can be readily applied to different energy-dependences of the transmission through the scatterer, like a monotonic increase in the case of a tunnel barrier or the multiple-resonant behavior associated with the excited states of a quantum dot. In addition, different generalizations of our model can be envisioned in order to describe other situations of experimental interest. This could include, for instance, finite-thickness wires and nanostructured two-dimensional electron gases.

\section*{Acknowledgments}
We gratefully acknowledge financial support from the French National Research Agency ANR through project ANR-20-CE30-0028 (TQT). This research was funded by the IKUR Strategy under the collaboration agreement between Ikerbasque Foundation and DIPC on behalf of the Department of Education of the Basque Government. We are extremely grateful to E.\ Zeldov for the stimulating discussions that encouraged us to carry out this research.
We thank H.\ Linke for discussions on transport in nanowires, and thank G.\ Percebois and G.\ Weick for interesting and fruitful discussions.

\appendix
\section{Solving the Boltzmann equation in strong Coulomb limit with momentum-independent relaxation time}
\label{appendix: Collision integral and Boltzmann}
In this appendix, we analyze the linearized Boltzmann equation in the convenient limit of strong Coulomb interaction and assuming momentum-independent relaxation time, $\tau(p)=\tau$, 
in the collision term in Eq.~\eqref{eq: collisionintegral}. 

Since the distribution~(\ref{eq: distribution function}) with the condition~(\ref{eq: potential}) gives the same density $n_0$ as the equilibrium distribution $f_0(p)$, the function $f_T(p,x)$ entering the collision integral is just $f_T(p,x)=f_0(p)$.
Then, plugging the distribution~(\ref{eq: distribution function}) into Eq.\ \eqref{eq:boltzmann} for an order-by-order solution yields two coupled inhomogeneous differential equations (one per branch $\pm p$) that depend on $e\varphi(x)$: 
\begin{align}
\label{appendix equation: homogenous correction, const. tau const density}
	\left[1\pm l(p)\,\partial_x  \right]\delta f(\pm p,x) =e\varphi(x)\partial_{\mu_0} f_0(p)\, ,
\end{align}
with the electron-phonon scattering length $l(p) = p\tau/m$. Using the self-consistency condition \eqref{eq: potential} and introducing the symmetric (s) and antisymmetric (a) combinations
\begin{align}
\label{appendix equations: symmetric/ asymmetric combinations}
	 \delta f^{\rm s, \rm a}(p,x) = \delta f(p,x) \pm\delta f(-p,x)\, ,
\end{align}
we arrive at a homogeneous second-order integro-differential equation for $\delta f^{\rm s}$,
\begin{align}
\label{appendix equation: diff equation for symmetric solution}
&	\left[1 - l^2(p)\,\partial^2_x\right]\delta f^{\rm s}(p,x) = \mathcal{C}(p) \int_0^\infty \frac{{\rm d}p'}{h}\,\delta f^{\rm s}(p',x),
\end{align}
with the normalized coefficient
\begin{align}
\label{appendix equation: coefficient C(p)}
	 \mathcal{C}(p) \equiv 2\,\frac{\partial_{\mu_0} f_0(p)}{\partial_{\mu_0} n_0},\quad
	 \int_0^\infty\frac{{\rm d}p}{h} \,\mathcal{C}(p) = 1 ,
\end{align}
and the relation between the symmetric and antisymmetric components
\begin{align}
\label{appendix equation: deltafa}
	 \delta f^{\rm a}(p,x) = -l(p)\,\partial_x \delta f^{\rm s}(p,x) \, .
\end{align}
Looking for solutions of Eq.~\eqref{appendix equation: diff equation for symmetric solution} with exponential spatial dependence $\delta f^{\rm s}(p,x)=g(p)\,{e}^{\pm\varkappa{x}}$, we transform Eq.\ \eqref{appendix equation: diff equation for symmetric solution} into a generalized eigenvalue problem
\begin{align}
\label{appendix equation: eigenfunction collison integral, const. tau const. denisty}
	\left[1-\varkappa^2\,l^2(p)\right] g(p) = \mathcal{C}(p)\,\int_0^\infty \frac{{\rm d}p''}{h}\, g(p'')\, .
\end{align}
Its kernel is separable, so all eigenfunctions are easily found. Apart from the zero mode $g(p)=\mathcal{C}(p)$ corresponding to the eigenvalue $\varkappa=0$, all others can be labeled by a momentum $p'>0$, so that $\varkappa_{p'}=1/l(p')=m/(p'\tau)$ and 
\begin{align}
\label{appendix equation: ansatz for g}
	g_{p'}(p) = h\,\delta(p-p') + Z_{p'}\, \mathcal{P}\, \frac{(p')^2\mathcal{C}(p)}{p^2-(p')^2}
\end{align}
with 
\begin{align}
Z_{p'} = -\left[ 1+ \mathcal{P} \int_0^\infty \frac{{\rm d}p}{h}\, \frac{(p')^2\mathcal{C}(p)}{p^2-(p')^2}\right]^{-1} \, ,
\end{align}
where $\mathcal{P}$ stands for the principal value of the corresponding integral. 
The first term on the r.h.s.\ in \eqref{appendix equation: ansatz for g} can be understood as an injected beam of particles with initial momentum $p'$ and the second term its inelastically scattered counterpart. 

Using the eigenfunctions $g_{p'}(p)$ as a basis, we expand the unknown $\delta f^{\rm s}(p,x)$ in this basis with coefficients $A_{\LL}(p')$, $A_\RL(p')$ for $x<0$ and $x>0$, respectively:
\begin{align}\label{appendix equation: ansatz for symmetric solution}
	\delta f^{\rm s}(p,x) = \tau\int_0^\infty \frac{\mathrm{d}p'}{h} \,A_{\LL,\RL}(p')\,g_{p'}(p)\, e^{-|x|/l(p')}.
\end{align}
The constant prefactor $\tau$ can be absorbed into $A_{\rm L,R}(p')$,
but we keep it separate for consistency with the general case in Appendix \ref{appendix generic case}. 
We include in Eq.~\eqref{appendix equation: ansatz for symmetric solution} only the exponentials that decay away from the scatterer.
Formally, the boundary condition at $x=\pm\infty$ also introduces terms that decay exponentially from the end of the wires at $x=\pm\infty$, but we drop these terms because they play no role at finite $x$.
Note that the non-decaying zero mode $\mathcal{C}(p)$ is not included in the expansion \eqref{appendix equation: ansatz for symmetric solution}. Even if it was included, it will eventually drop out of the distribution function, canceled by a constant shift in $e\varphi(x)$ by virtue of Eq.~\eqref{eq: potential}.

With the form \eqref{appendix equation: ansatz for symmetric solution}, Eq.~\eqref{eq: potential} for $e\varphi(x)$ can be expressed in terms of the contributions arising from the different $p'$ modes as  
\begin{align}
e\varphi(x)=\frac{\tau}{\partial_{\mu_0} n_0}
\int_0^\infty \frac{{\rm d}p'}{h} A_{\LL,\RL}(p') e^{-|x|/l(p')}
\int_0^\infty \frac{{\rm d}p}{h} g_{p'}(p).
\end{align}
The current conservation is ensured by the fact that the corrections $\delta f$ do not contribute to the current \eqref{eq: current}, which follows from the relation
\begin{align}
\label{appendix equation: particle number conservation for eigenfunctions}
 \int_0^\infty \frac{{\rm d}p}{h}\,p^2\,g_{p'}(p) = 0\, ,
\end{align}
that is fulfilled by the solutions \eqref{appendix equation: ansatz for g}.

Going back to the equations \eqref{appendix equations: symmetric/ asymmetric combinations}, the form \eqref{appendix equation: ansatz for symmetric solution} of $\delta f^{\rm s}(p,x)$ and its relation \eqref{appendix equation: deltafa} with $\delta f^{\rm a}(p,x)$,
we can write the full distribution functions \eqref{eq: distribution function} including all correction terms as
\begin{subequations}
\label{appendix equation: distributions}
\begin{align}    
	f(p,x<0) =& f_0(p) + f_{\rm D}(p) \notag \\  
	&+\frac{\tau \mathcal{C}(p)}{2}\int_0^\infty \frac{\mathrm{d}p'}{h}\,\frac{pZ_{p'}A_\LL(p')}{p+{p'}} e^{x/l(p')},
	\label{appendix equation: left wire "inflow"}\\	
	f(-p,x>0) =& f_0(p) - f_{\rm D}(p) \notag \\
	&+ \frac{\tau \mathcal{C}(p)}{2}\int_0^\infty \frac{\mathrm{d}p'}{h}\,\frac{pZ_{p'}A_\RL(p')}{p+{p'}} e^{-x/l(p')},
	\label{appendix equation: right wire "inflow"}\\
	f(-p,x<0) =& f_0(p) - f_{\rm D}(p) + \tau A_\LL(p)\, e^{x/l(p)} \notag \\
	&+ \frac{\tau \mathcal{C}(p)}{2}{\mathcal{P}}\int_0^\infty \frac{\mathrm{d}p'}{h}\,\frac{pZ_{p'}A_\LL(p')}{p-{p'}} e^{x/l(p')},
	\label{appendix equation: left wire "outflow"}\\
	f(p,x>0) =& f_0(p) + f_{\rm D}(p) + \tau A_\RL(p)\, e^{-x/l(p)} \notag \\
	&+ \frac{\tau \mathcal{C}(p)}{2}{\mathcal{P}}\int_0^\infty \frac{\mathrm{d}p'}{h}\,\frac{pZ_{p'}A_\RL(p')}{p-{p'}} e^{-x/l(p')}.
	\label{appendix equation: right wire "outflow"}
\end{align}
\end{subequations}
The first two expressions \eqref{appendix equation: left wire "inflow"} and \eqref{appendix equation: right wire "inflow"} describe the right (left) movers that approach the scatterer from the left (right) sector of the wire, while the last two expressions \eqref{appendix equation: left wire "outflow"} and \eqref{appendix equation: right wire "outflow"} describe the left (right) movers that move away from the scatterer in the left (right) sector and thus contain the additional term $A_{\LL,\RL}(p)\, e^{\pm x/l(p)}$ representing particles injected through the scatterer. 

To conclude this appendix, we give the derivation of Eq.~\eqref{eq:Ps=phi}. From the definition, Eqs.~\eqref{eq: PQPC}, we have
\begin{align}
   \Pphon^\text{s}(x) {}&{} = -\int_0^\infty\frac{\mathrm{d}p}{h}\,\frac{p^2}{2m}\,\frac{p}{m}\,\frac{\partial\delta f^\mathrm{a}(p,x)}{\partial{x}}\nonumber\\ {}&{}= \int_0^\infty\frac{\mathrm{d}p}{h}\,\frac{p^2}{2m}\,\frac{p^2\tau}{m^2}\,\frac{\partial^2\delta f^\mathrm{s}(p,x)}{\partial{x}^2}\nonumber\\{}&{}= \int\limits_0^\infty\frac{\mathrm{d}p}{h}\,\frac{p^2}{2m\tau}\left[\delta f^\mathrm{s}(p,x)-\mathcal{C}(p) \int\limits_0^\infty \frac{{\rm d}p'}{h}\,\delta f^\mathrm{s}(p',x)\right].
\end{align}
The first term in the square brackets drops out by virtue of Eq.~\eqref{appendix equation: particle number conservation for eigenfunctions}. For the second term we use Eq.~\eqref{eq: potential}, which gives
\begin{align}
   \Pphon^\text{s}(x) {}&{} = -\frac{e\varphi(x)}{2\tau}\,2\int_0^\infty\frac{\mathrm{d}p}{h}\,\frac{p^2}{m}\,\frac{\partial{f}_0(p)}{\partial\mu_0}\nonumber\\ {}&{} = \frac{e\varphi(x)}{2\tau}\,2\int_0^\infty\frac{\mathrm{d}p}{h}\,p\,\frac{\partial{f}_0(p)}{\partial{p}}
    = -\frac{n_0}{2\tau}\,e\varphi(x).
\end{align}
\section{Momentum discretization and matching conditions}
\label{appendix: matching}
This appendix highlights some subtleties of the numerical solution of the model explained in Sec.~\ref{section: model}. Specifically, we outline how we discretize the problem so that it reduces to a finite (but large) number of coupled linear equations that can be solved numerically.

The unknown expansion coefficients $A_{\LL,\RL}(p')$ determining the distribution function via Eqs.~\eqref{appendix equation: distributions} should be found from the boundary conditions at $x=0^\pm$. Namely, plugging Eqs.~\eqref{appendix equation: distributions} into the matching conditions in Eq.~\eqref{eqs: matching condition}, we obtain a system of linear equations for $A_{\LL,\RL}(p')$. To solve them numerically, we have to discretize the momenta in order to work with a finite set of equations. 
We start with some initial guess for the voltage drop across the scatterer, $eV>0$. Thus, we choose a discrete grid in energy, which induces two different grids for the momenta on the left and on the right via Eq.~\eqref{eq: kinetic energies}, $p_\LL^i$ and $p_\RL^j$ with $i=1,\ldots,N_\LL$, $j=1,\ldots,N_\RL$, $N_\RL>N_\LL$. 

The solution of the discretized version of the eigenvalue problem~\eqref{appendix equation: eigenfunction collison integral, const. tau const. denisty} produces $N_\LL$ ($N_\RL$) eigenvectors on the left (right) side of the scatterer. On each side, there is one eigenvector corresponding to the zero eigenvalue $\varkappa=0$ (the zero mode). As mentioned in Appendix~\ref{appendix: Collision integral and Boltzmann}, the coefficients of the zero modes drop out of the distribution functions \eqref{appendix equation: distributions} by virtue of Eq.~\eqref{eq: potential}. Thus, we have $N_\LL+N_\RL-2$ unknown coefficients to determine from the matching conditions.

The matching equations~\eqref{eqs: matching condition}, when put on the discrete energy grid, produce $2N_\LL$ equations for $\epsilon>eV$, as well as $N_\RL-N_\LL$ additional equations for $0<\epsilon<eV$ imposing total reflection for particles incident from the right. Thus, we have $N_\LL+N_\RL$ equations for $N_\LL+N_\RL-2$ unknown coefficients. 

However, Eqs.~\eqref{eqs: matching condition} are not linearly independent. Indeed, Eq.~\eqref{eq:IQPCright} is a linear combination of equations obtained from Eq.~\eqref{eq:matching_right} on the grid, and the same holds for Eq.~\eqref{eq:IQPCleft} and Eq.~\eqref{eq:matching_left}.
Due to the relation~\eqref{appendix equation: particle number conservation for eigenfunctions}, all unknown coefficients drop out from the right-hand side of Eqs.~\eqref{eq:IQPC} which amounts to~$I$ for both equations. Thus, two seemingly different equations \eqref{eq:IQPCright} and \eqref{eq:IQPCleft} are in fact identical.

As a result, we have $N_\LL+N_\RL-1$ independent linear equations for $N_\LL+N_\RL-2$ unknown coefficients. They can be compatible only if an additional condition is imposed. Since their very construction requires us to assume a value of $eV$, while only $\Fext$ enters the constant term of the equations, it is easier to determine $\Fext$ for given $eV$. To then find $eV$ for a given $I$, we have to iterate the self-consistency loop until convergence.
\section{Current-voltage relation}
\label{appendix: I-eV relation}
\begin{figure}[t!]
	\centering
	\includegraphics[width = 0.9\columnwidth]{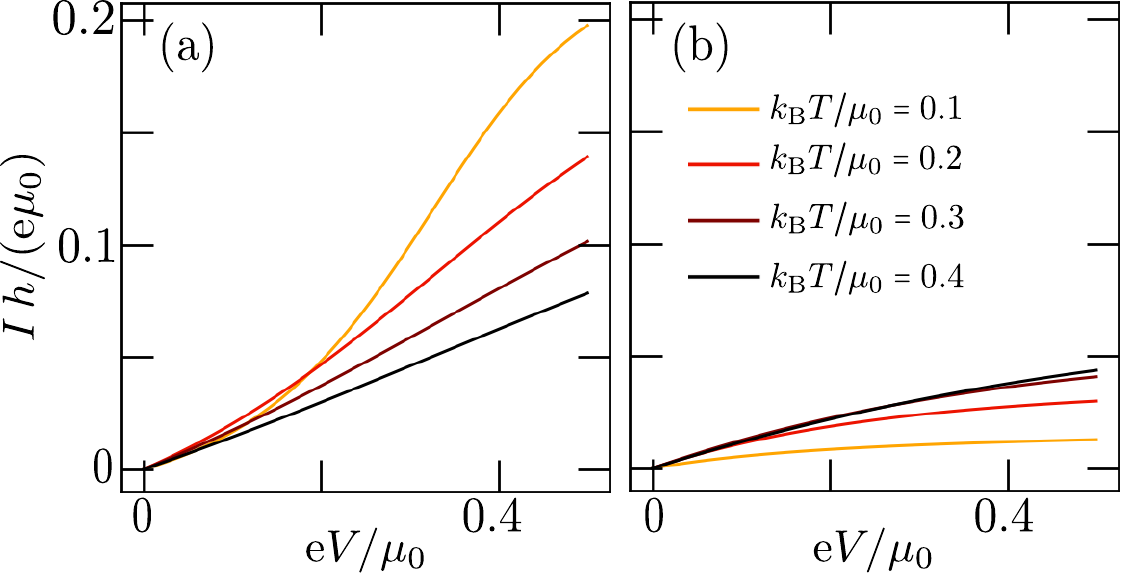}
	\caption{Current-voltage ($I$-$V$) curves for a Lorentzian transmission with $\Gamma/\mu_0 = 0.08$ at different temperatures. (a) For $\epsilon_0/\mu_0 = 0.64$, larger temperature increases the resistivity. (b) The opposite tendency is observed when the transmission peak is at $\epsilon_0/\mu_0 = 1.44$.}
	\label{Fig-IV-relation}
\end{figure}

The numerical approach described in Sec.~\ref{section: model} was devised to determine the phonon heating in the wires, under the assumption of strong Coulomb repulsion and momentum-independent relaxation time. As a byproduct, the numerics also gives the self-consistent current-voltage characteristics.
Such a relation is of interest, particularly in the case of arbitrary temperature and  voltage drop across the scatterer.

Fig.~\ref{Fig-IV-relation} shows this current-voltage relation for the Lorentzian transmission of Eq.~\eqref{eq: Lorentzian transmission} with $\Gamma/\mu_0 = 0.08$ at different temperatures. The voltage we chose to plot here is the voltage drop across the scatterer itself, $V$, rather than the voltage drop in the vicinity of the scatterer, $V_{\rm tot}$ (see Fig.~\ref{Fig_sketch_of_potential}). The current-voltage relation is generally non-linear and strongly temperature-dependent. An increase in temperature tends to increase the current when the transmission is peaked above the Fermi energy at $x=0^-$, since a thermal population of particles on the left is needed for the current to flow through the transmission window. For the transmission peaked below the Fermi level on the left, the transmission window has a strong overlap with the energy range where where $f_0(\epsilon-eV)-f_0(\epsilon)$ is significant. Increasing temperature then leads to a decrease in the current due to a reduction of $f_0(\epsilon-eV)-f_0(\epsilon)$ at energies where the transmission is large, accompanied by an increase in energy ranges where the transmission is small. In all cases the current has a tendency to saturate at high voltages, since it has a natural upper bound of $(e/h)\int_{-\infty}^\infty\mathcal{T}(\epsilon)\,d\epsilon$ [see Eqs.~(\ref{eq:IQPC})] when the integral converges.
\section{Boltzmann equation in general case of finite Coulomb repulsion with momentum-dependent relaxation}
\label{appendix generic case}
While the numerical results discussed in the main text refer to the case of a momentum-independent relaxation time in the strong Coulomb limit, the approach to obtain the phonon heating is rather general. We thus present in this appendix the generalization of the case with a momentum-dependent relaxation time and finite Coulomb repulsion. 

The case without scatterer (\textit{i.e.}\ $\Transmission \equiv 1$) is particularly simple to generalize following the approach of Appendix~\ref{appendix: Collision integral and Boltzmann} since $e\varphi(x)\equiv 0$ and no charge is accumulated within the wires. In particular, we still have $f_T(p,x) = f_0(p)$ with the constant chemical potential~$\mu_0$, implying that $f (\pm p,x) = f_0(p) \pm f_{\rm D}(p)$, with $f_{\rm D}(p)$ given by Eq.~\eqref{eq: Drude term}. This results in a constant charge density $n(x)=n_0$.

The presence of the scatterer with a finite reflection leads to charge accumulation. We thus write 
\begin{subequations}
\begin{align}
	n(x) &= n_0 +\delta n(x),\\
	f_T(x) &=\left[e^{\beta[p^2/(2m) - \mu_0 -\delta \mu(x)]}+1\right]^{-1} \nonumber\\ 
 &\approx  f_0(p) + \delta \mu(x) \,\partial_{\mu_0}f_0(p).
 \end{align}\end{subequations}
Always working with the linearized kinetic equation, we only need to consider corrections $\delta n(x)$ and $\delta \mu(x)$ of order $\Fext$. The scheme introduced in the beginning of Sec.\ \ref{section: model} [Eqs.\ \eqref{eq:boltzmann}--\eqref{eq: total density}] is general, and in particular, the ansatz~\eqref{eq: distribution function} can be used for the solution of the linearized case. The possibility of having charge accumulation implies
\begin{align}
\label{appendix equation: particle density, generic case}
	 \delta n(x) = -e\varphi(x) \frac{\partial n_0}{\partial \mu_0} + \int_0^\infty \frac{{\rm d}p}{h} \left[\delta f(p,x)+\delta f(-p,x)\right].
\end{align}
The condition of the particle conservation by the collision integral \eqref{eq: definition mu(x)} now translates into
\begin{subequations}
\begin{align}\label{appendix equation: particle number conservation, coll. integral, lowest order}
	&\delta \mu(x) + e\varphi(x) = \kappa\,\int_0^\infty \frac{{\rm d}p}{h}\,\frac{\delta f(p,x)+\delta f(-p,x)}{\tau(p)},\\
 &\kappa\equiv\left[2\int_0^\infty\frac{{\rm d}p}h\,\frac{\partial_{\mu_0}f_0(p)}{\tau(p)}\right]^{-1}.
\end{align}\end{subequations}
The corrections $\delta f(\pm p,x)$ of the ansatz \eqref{eq: distribution function} satisfy a generalization of Eq.\ \eqref{appendix equation: homogenous correction, const. tau const density} that reads 
\begin{align}\label{appendix equation: homogenous correction}
	\left[1\pm l(p)\,\partial_x \right]\delta f(\pm p,x) = [\delta\mu(x)+ e\varphi(x)]\partial_{\mu_0}f_0(p)\, ,
\end{align}
where we now use the definition 
\begin{align}
l(p) = \tau(p)p/m
\end{align}
of the electron-phonon scattering length.

Such a system of differential equations can be conveniently handled by using the 
symmetric (antisymmetric) combinations $\delta f^\mathrm{s,a}(\pm p,x)$ defined in Eq.\ \eqref{appendix equations: symmetric/ asymmetric combinations}. 
We can then trade the previous system by a second-order equation for the symmetric combination  
\begin{align}
	\left[1- l^2(p)\partial^2_x \right]\delta f_{\rm s}(p,x) = 2[\delta\mu(x)+ e\varphi(x)]\partial_{\mu_0}f_0(p)\, ,
\end{align}
together with Eq.\ \eqref{appendix equation: deltafa}.
Proceeding along the lines of Appendix \ref{appendix: Collision integral and Boltzmann}, the use of Eq.\ \eqref{appendix equation: particle number conservation, coll. integral, lowest order} allows us to write the previous differential equation as  
\begin{align}\label{appendix equation: diff. equation for symmetric solution, generic case}
	\left[1- l^2(p)\partial^2_x \right]\frac{\delta f^\mathrm{s}(p,x)}{\tau(p)} = \mathcal{C}_\tau (p) \int\limits_0^\infty \frac{{\rm d}p'}{h}\,\frac{\delta f^\mathrm{s}(p',x)}{\tau(p')}\, ,
\end{align}
with the normalized and $\tau(p)$ dependent coefficient 
\begin{align}
	\mathcal{C}_\tau (p) \equiv \frac{2\kappa}{\tau(p)}\,\partial_{\mu_0} f_0(p), \quad \int_0^\infty \frac{{\rm d}p}{h}\,\mathcal{C}_\tau (p) = 1.
\end{align}
Notice that $\mathcal{C}_\tau (p)$ reduces back to $\mathcal{C}(p)$ from Eq. \eqref{appendix equation: coefficient C(p)} for a constant $\tau(p)$.
Looking for solutions of Eq.~\eqref{appendix equation: diff. equation for symmetric solution, generic case} in the form
\begin{equation}
    \frac{\delta f^\mathrm{s}(p,x)}{\tau(p)} = g(p)\,e^{\pm\varkappa{x}},
\end{equation}
we arrive at the eigenvalue equation which differs from Eq.~\eqref{appendix equation: eigenfunction collison integral, const. tau const. denisty} by the replacement $\mathcal{C}(p)\to\mathcal{C}_\tau(p)$. Again, it has the zero mode eigenfunction, $g(p)=\mathcal{C}_\tau(p)$, while other eigenfunctions can be labeled by a momentum $p'>0$, so that $\varkappa_{p'}=1/l(p')$:
\begin{subequations}
\begin{align}\label{appendix equation: generic eigenfunctions}
&	g_{p'}(p)  = h\,\delta (p-p') + Z_{p'}\,{\mathcal{P}} \frac{l^2(p')\,\mathcal{C}_\tau(p)}{l^2(p)-l^2(p')},\\
&   Z_{p'} = -\left[1+\mathcal{P}\int_0^\infty\frac{\text{d}p}h\,\frac{l^2(p')\,\mathcal{C}_\tau(p)}{l^2(p)-l^2(p')}\right]^{-1}.
\end{align}
\end{subequations}
The current conservation is ensured by the property 
\begin{align}
 \int_0^\infty \frac{{\rm d}p}{h}\,l^2(p)\,g_{p'}(p) = 0.
\end{align}
Instead of Eq.~\eqref{appendix equation: ansatz for symmetric solution}, the expansion in eigenfunctions has the form
\begin{align}
	\delta f^{\rm s}(p,x) = \tau(p)\int\frac{\text{d}p'}h \,A_{\LL,\RL}(p')\,g_{p'}(p)\, e^{-|x|/l(p')},
\label{appendix equation: correction from Boltzmann, generic case}
\end{align}
with the coefficients $A_{\LL}(p')$ and $A_{\RL}(p')$ taken for $x<0$ and $x>0$, respectively, to be found from the matching equations at the scatterer. Then, the self-consistency loop involving Poisson's equation and Eq.~\eqref{appendix equation: particle density, generic case} should be converged. Once this convergence is achieved, $\delta \mu(x)$ can be deduced from Eq. \eqref{appendix equation: particle number conservation, coll. integral, lowest order} as a byproduct. By virtue of Eqs.~\eqref{eq: PQPC}, the phonon heating is given by a combination of many different modes that decay at different rates with increasing $|x|$. Just like for the special case analyzed in the body of this work, we expect that the general case will commonly have some decaying modes with positive amplitudes and other decaying modes with negative amplitudes.
Then it is likely to exhibit heating and cooling spots
peaked at a finite distance from the scatterer, in a similar manner to the special case studied in the body of this work.

There is, however, a notable exception to the above statements: the case  $\tau(p)\propto 1/p$, leading to $l(p)=\mathrm{const}$.  This case was studied in Ref.~\cite{eranen1987}, showing that it allows an exact solution.
There, all eigenfunctions have the same eigenvalue $\varkappa=1/l$, apart from the zero mode eigenfunction, $g(p)=\mathcal{C}_\tau(p)$.  This massively degenerate subspace is defined by the condition $\int_0^\infty{g}(p)\,\text{d}p=0$, and the functional basis in this subspace can be chosen arbitrarily. In particular, it can be chosen to resemble Eq.~(\ref{appendix equation: generic eigenfunctions}), namely, $g_{p'}(p)=h\,\delta(p-p')-\mathcal{C}_\tau(p)$.
The crucial feature of the case $l(p)=\text{const}$ is that the spatial dependence $e^{-|x|/l}$ factorizes out, 
and the 
$x$-dependence of $\delta f^s(p,x)$ in Eq.~(\ref{appendix equation: correction from Boltzmann, generic case})
becomes the same exponential decay for all $p$,
so the explicit form of the eigenfunctions becomes effectively unimportant for this exponential decay. This is the most obvious situation demonstrating that the relaxation time $\tau(p)$ is the crucial parameter of the model.

We can use the exact solution  for  $l(p)=l=\text{const}$ \cite{eranen1987} to look in detail at the spatial dependence of phonon heating in this case. As $\delta f^s(p,x)$ in Eq.~(\ref{appendix equation: correction from Boltzmann, generic case}) has the same decay with distance $x$ for all $p$, we know that $\delta f^s(p,x)$ itself must decay monotonically with $x$, so $\Pphon^{\rm s}(x)$ alone cannot lead to heating or cooling spots.  In contrast, outside the strong Coulomb limit, $e\varphi(x)$ exhibits decay on another length scale, the screening length.
Thus, the combined effect of $\Pphon^{\rm s}(x)$ and $\Pphon^{(\varphi)} (x)$ may still lead to heating or cooling spots.

\bibliography{ref}

\end{document}